# Information and the living tree of life

A theory of measurement grounded in biology

**Kevin Hudnall**
kahudnall@ucdavis.edu
University of California, Davis
Biological Systems Engineering Graduate Group
One Shields Avenue, Davis CA 95616 USA

## Highlights

- Extends prior work deriving time from life's structure to a theory of measurement
- Develops an information-theoretic framework unifying time and observation
- Introduces a dilation equation making observer-relative time computable
- Establishes *biokinematics*: a framework for measuring life in motion across species
- Shows standards like the SI second are biologically dependent but still operational

## Abstract

We extend a formal framework that previously derived time from the multifractal structure of biological lineages (Hudnall & D'Souza, 2025). That work showed that time itself is multifractal – not a universal background dimension, but an observer-dependent geometry. Here we develop the corresponding theory of measurement: showing that a multifractal conception of time not only permits measurement, but grounds it more rigorously in the structure of biology. The tree of life is modeled as the outcome of stochastic, convex branching, and we show how information-theoretic and fractal measures render its multifractal geometry into measurable, observer-relative time intervals. At the core is a dilation equation that expresses relative time elapse between entities as dimensionless ratios. Operational standards such as the SI second remain valid, but our framework makes explicit their lineage-dependence. This framework unifies measurement theory with biological form, preserves full compatibility with established science, and provides a biologically grounded theory of observation. It enables comparative analyses of duration and kinematics across lineages, with predictions that are directly open to experimental validation.

## Keywords

# 1 Introduction

Measurement is one of the most fundamental activities in science, and time stands at its foundation. In the SI system, virtually all other operational definitions trace back to the second. In practice, time is employed as a coordinate, made variable through relativity, but assumed to be non-variable with respect to the biology of measurement itself. Biology, however, has long recognized that the pace of life varies across species – from circadian rhythms (Dunlap, 1999) to metabolic rates (West et al., 1997) to life-history schedules (Charnov, 1993). These observations, together with broader evolutionary scaling patterns such as heterochrony (McKinney & McNamara, 1991) and molecular rate variation (Bromham, 2011), suggest that biology may have more to say about the nature of time than is often acknowledged. Yet measurement in biology typically follows the rest of science, assuming time as an external backdrop in which life unfolds, without fully recognizing the biological dependence of all measurement.

This gap between biology's recognition of lineage-dependent variability and science's assumption of a universal backdrop defines the problem this work addresses. While models in both physics and biology capture aspects of temporal variability, none provide a lineage-grounded account of how measurement itself becomes possible. Recent work of Hudnall & D'Souza (2025) introduced a formal model showing that the tree of life, modeled as a process of stochastic convex branching, generates a multifractal geometry from which time emerges as a lineage-dependent observable. This makes time not a universal background dimension, but a multifractal geometry with as many dimensions as there are living things. That paper established the base geometry. The present paper is fully self-contained and builds upon that foundation to develop the corresponding theory of measurement.

Here we demonstrate that a multifractal conception of time not only permits measurement, but grounds it within a more rigorous, lineage-dependent geometry. We show how the multifractal structure can be operationalized using information-theoretic and fractal quantities, leading to a dilation equation that quantifies relative time intervals as dimensionless ratios. In this view, operational standards such as the SI second remain valid, but their lineage-dependence is made explicit, enabling cross-species conversion through the shared information geometry of descent.

This extension renders the theory of multifractal time empirically tractable. Branch lengths retain their phylogenetic meaning, but no single global metric is assumed across the tree. Instead, observer-relative nearness becomes measurable once centered on a specific organism, allowing durations to be expressed in fully operationalized units such as the SI second. In this way, the complete measurement system of modern science is retained, but now explicitly grounded in the biological structure of descent, formalizing what has long remained implicit in practice.

This paper formalizes that structure. The remainder of Section 1 summarizes the biological foundations of multifractal time from (Hudnall & D'Souza, 2025). Section 2 formalizes the geometry and information structure, culminating in a lineage-dependent dilation equation for observer-relative time. Section 3 presents simulations, computes key information quantities, and demonstrates how durations can be recovered in standard units, yielding experimental tractability. Section 4 reframes measurement as a biological act emergent from, and made possible by, the informational structure of life.

## 1.1 The biological foundations of observer-dependent time

The multifractal structure of time emerges from the interplay of three foundational principles that are both well-established and widely employed – nestedness, duality, and randomness (Hudnall & D'Souza, 2025). While familiar in themselves, their integration here is formalized within an information-theoretic structure that links observation, measurement, and lineage.

The *nested principle* captures the inherent convexity of descent: the probability of a descendant is necessarily less than that of its ancestor. Each lineage contracts in probability space, forming a sequence of increasingly constrained distributions and a unique series of nested partitions determined by its ancestry (Appendix A.1).

The *duality principle* frames reproduction as fission: a single biological form gives rise to a random number of descendants, each with uncertain but constrained form (Appendix A.2). These convex branching events shift topological dimension – from point-like entities of dimension zero to tree-like populations of dimension one – generating a dynamic "trees-within-trees" hierarchy consistent with modern phylogenetic theory (Page & Charleston, 1998).

The *random principle* renders the other two mathematically precise via probability functions (Appendix A.3, Definition A2), capturing the stochastic nature of reproduction and yielding an exact definition of a living entity as a random variable with nontrivial probability of change. Defining the living is crucial for linking biological structure to measurement, since only living entities qualify as observers (Section 2.3). The living tree of life is modeled as a random iterated function system (Appendix B), with each fission event introducing probabilistic variation in scale and structure. As a result, each lineage is a unique realization of a random process with nonpositive entropy that increases in absolute value over generative steps (Section 2.6.1).

In this framework, each lineage is a probabilistically convex trajectory through the tree, corresponding to a distinct monofractal set (Appendix C.1) shaped by its unique sequence of events. **Figure 1** shows the principles in combination, producing the living tree of life as a dynamic structure.

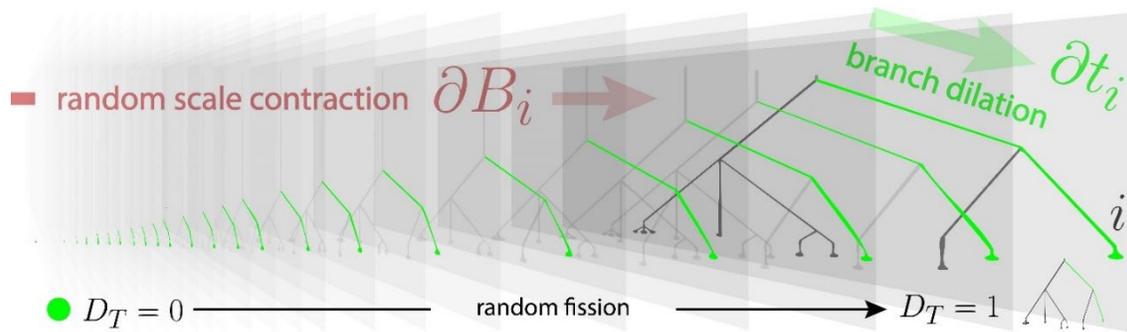

**Figure 1** *The living tree of life emerges from three principles – nestedness, duality, and randomness.* Each reproduction event is a random fission, shifting the topological dimension $D_T$ from 0 to 1. These fissions take the form of nested contractions of random biological variables $\partial B_i$, with corresponding branch dilations ($\partial t_i$) as lifeforms unfold. This framework renders the tree of life dynamic – a system in which $\partial B_i/\partial t_i \equiv (\partial B/\partial t)_i$ exist and are nonzero. While lifespans are shown, temporal structure is not presupposed; see Section 1.5 for details.

These principles do not presuppose time; rather, by formalizing them mathematically, time itself emerges as a multifractal, lineage-dependent observable. In this way, the principles are directly tied to time: it is their mathematical integration that generates time as an outcome. This generative framework sets the stage for deriving time as an emergent property of the living tree of life – a quantity defined not a priori, but through the probabilistic geometry of lineages.

## 1.2 An information structure

These branching trajectories carry not just structural form but informational content. Each lineage traces a unique path through probability space, producing a distinct informational signature. In our simulation (**Animation 1**), two example lineages are visualized, each generating its own sequence of random variables, $B_i$ and $B_j$. We track their entropies $H[B_i]$, $H[B_j]$, their joint entropy $H[B_i, B_j]$, mutual information $I[B_i; B_j]$, and information distance $d_\beta(B_i, B_j)$, with formal definitions given in Section 2.6.

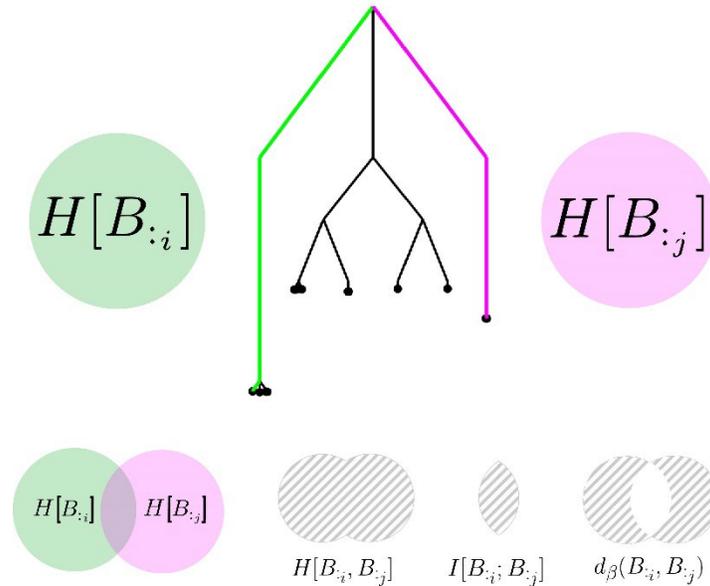

**Animation 1** *Path entropies in the multifractal tree of life.* Two lineages (green and purple) trace nested sequences of random variables through the branching tree.

This information structure is the basis for quantifying time intervals on the tree while respecting its multifractality (Section 2.7). It also distinguishes between the living – those still undergoing fission – and the dead, where fission has ceased. Since only the living exhibit ongoing evolutionary dynamics, this distinction is fundamental to any dynamic theory of phylogeny.

## 1.3 Life as approximation to the present

This formalism leads directly to a dynamic definition of the present. Life is defined by the capacity for change (Section 2.3): a biological entity is alive if its future form remains uncertain, and dead once fixed. The living portion of the tree of life – the set of all lineages still changing – forms a moving boundary: the present. Each reproductive event redefines this boundary, pushing what is living now into the past. In this sense, *the living define the present*.

This parallels continuous phase-transition models in which biological systems persist at criticality (Longo & Montévil, 2014). Because living entities are always in motion, the present can only be approximated: once a lineage is observed, it has already changed. Our model formalizes this asymmetry by treating the living as the unresolved limit of an ongoing process (Sections 2.3.2).

Approximation here is not a weakness but a defining property of life – one that shapes the simulation framework developed in Section 3.

## 1.4 Time as an emergent multifractal geometry

Recursive branching generates a multifractal geometry (Section 2.4): each lineage is a monofractal curve shaped by its ancestral constraints. Collectively, these curves make the tree of life multifractal: a set of lineage-specific fractals, each with its own scaling behavior. In such a structure, temporal distances depend on biological form – time becomes *observer-relative*. **Animation 2** visualizes this dependence colloquially, contrasting two lineages with different ancestral histories, showing that the experienced duration between the same events can differ.

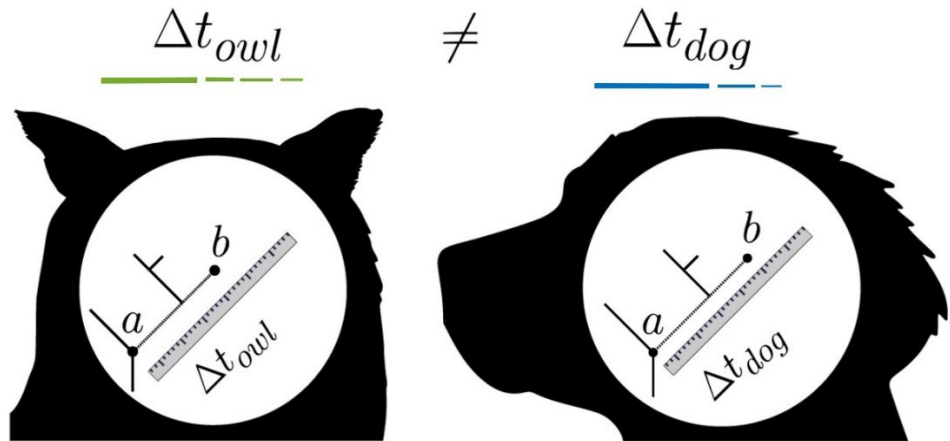

**Animation 2** *Multifractal time as lineage-specific dilation.* Two lineages – an owl (green) and a dog (blue) – trace distinct fractal trajectories through the tree. Asymmetric scaling causes the experienced duration between events $a$ and $b$ to differ: $\Delta t_{owl} \neq \Delta t_{dog}$. Common rulers indicate a shared physical standard (e.g., the cesium resonance frequency) on which all observers agree.

Information quantities (Section 2.6) make this observer-relative time measurable. Each lineage accumulates entropy as reproduction narrows its possible futures, and divergence between lineages is quantified via mutual information and information distance. These measures express elapsed time between paths not in absolute units, but as dimensionless ratios derived from the tree's own structure (Section 2.7). This is illustrated colloquially in **Figure 2**. Two observers trace their ancestry to a shared ancestor, but multifractal asymmetry means each experiences a different duration along its path, expressed as the dilation value $\partial t_j^i$ (Equation 18).

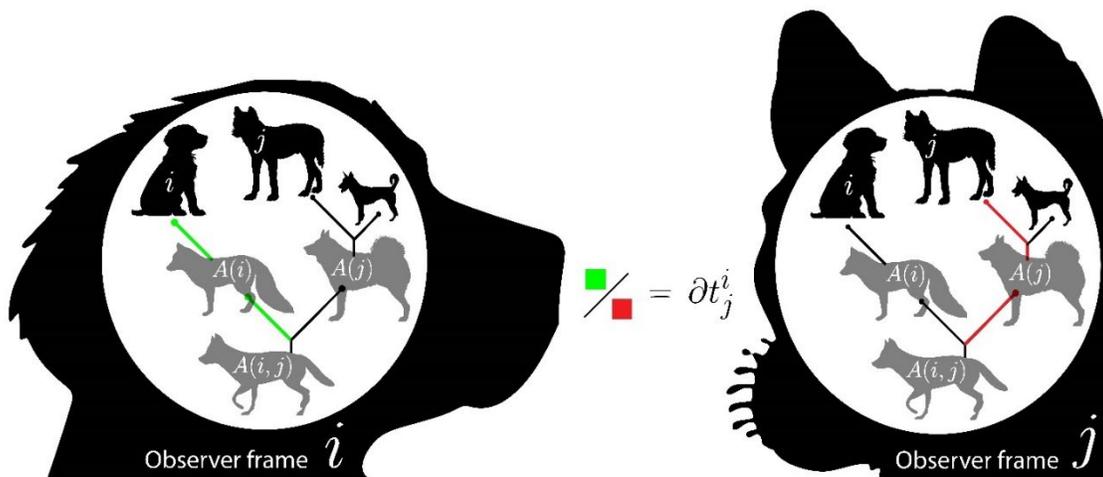

**Figure 2** *Relative time elapse as a lineage-dependent dilation ratio.* Two observers, $i$ and $j$, trace their paths to a shared ancestor $A(i,j)$. Each experiences a different duration, quantified by the dimensionless ratio $\partial t_j^i$.

Before formalizing this dilation, we clarify why the framework does not assume temporal structure.

## 1.5 No presupposition of time

The $\partial t_j^i$ illustrated in Figure 2 represent ratios of nearness, not metric durations. Our construction assumes lifespans in the structural sense – intervals bounded by birth and reproduction/death, indexed by reproduction events $n$ – but it does not assign these intervals a metric length. Prior to centering (Section 2.7.4), the living tree is non-metric. Instead, it is a nearness space in which order, inclusion, and proximity are defined but absolute distances are not (Appendix F).

Because lifespans lack assigned distances, the framework does not presuppose time as a universal background dimension. That is why $t$ in Figure 1 requires a unique subscript. Duration is later constructed from within the system using these dimensionless ratios of nearness derived from the information geometry. Only after centering on an observer does a local measure space arise, allowing operational units such as seconds to enter and nearness to become measurable time (Section 3.5).

## 1.6 From observation to measurement

Once time is constructed in this way, centering on an observer turns it into a measurable quantity. An observer is a living entity: an approximate endpoint of evolutionary descent that retains the capacity for change (Section 2.3). Its inherited constraints define a unique informational fingerprint (Section 2.6.1), enabling cross-lineage comparisons (Section 2.6.2). Centering the tree on a particular observer recovers a measure space: a reference frame in which durations become measurable (Section 3.5). Nearness ratios then act as conversion factors, translating durations of other entities into the temporal frame of the centered observer. Operational time units (such as the SI second) are thus observer-specific – meaningful only when tied to the lineage that defines them – yet remain translatable across lineages through the shared information geometry (Section 2.7).

Throughout this work, *time* refers exclusively to the quantity illustrated in Figure 2: the experienced durations of biological entities as determined by their relative positions in the tree.

# 2 Materials and methods

This section formalizes the model architecture and computational procedures for simulating the tree of life as a nested, dualistic, and probabilistic system (Hudnall & D'Souza, 2025), and develops the framework by which this structure becomes measurable. We begin by defining individual lineages as stochastic processes over ancestral paths (Section 2.1) and then describe the branching dynamics that generate the full tree structure (Section 2.2). Observers are defined as the living portion of the tree (Section 2.3), and the resulting geometry is characterized using fractal dimensions (Section 2.4). We then introduce the information-theoretic framework needed to quantify lineage relationships (Section 2.6), culminating in the dilation equation that expresses multifractal time intervals as dimensionless ratios (Section 2.7). Together, these steps provide a unified mathematical foundation for treating time as an emergent, observer-relative quantity, and for rendering it measurable.

## 2.1 Paths as nested stochastic processes

Each lineage is represented as a stochastic process — a sequence of random variables whose structure is shaped by the convex constraints of its ancestry. In this notation, individual biological events are denoted by lowercase $\beta$, with nested subscripts tracking the path back through ancestral events. The corresponding random variables are written in uppercase $B$, parameterized by reproduction events $n$ and ancestral form $\beta$.

**Table 1** and **Figure 3** serve as a quick reference for this indexing system: Table 1 defines the symbols, and Figure 3 visualizes how nested paths specify a unique biological trajectory through the tree of life. For clarity, a full table of symbols and notation used throughout the manuscript is provided in Appendix H.

**Table 1.** *Random variables of biological form and nested path notation.*

| Expression | Meaning | Notes |
|---|---|---|
| $\beta_{0:k_l}$ | Biological event $l$ in a lineage. | $\beta_0$ is the ancestor of all life; the colon represents an unspecified gap; $l$ is a descendent of $k$. |
| $\beta_{:l}$ or $\beta_l$ | Biological event $l$. | Shorthand omitting explicit ancestral indices. |
| $B_{:l_m}$ | Random variable $m$. | A function of reproduction events $n$, and its ancestor $\beta_{:l}$ |
| $A\left(B_{:l_m}\right)$ | Ancestor of $m$. | $A$ returns the most recent ancestor: $A\left(B_{:l_m}\right) = \beta_{:l}$. |
| $\mathbb{T}_n^{\text{Dead}}$ | The dead tree of life | Subset of all biological forms indexed by $n$ (Definition A1). |
| $\mathbb{T}_N^{\text{Alive}}$ | The living tree of life | All random variables with nontrivial probability (Definition A2). |

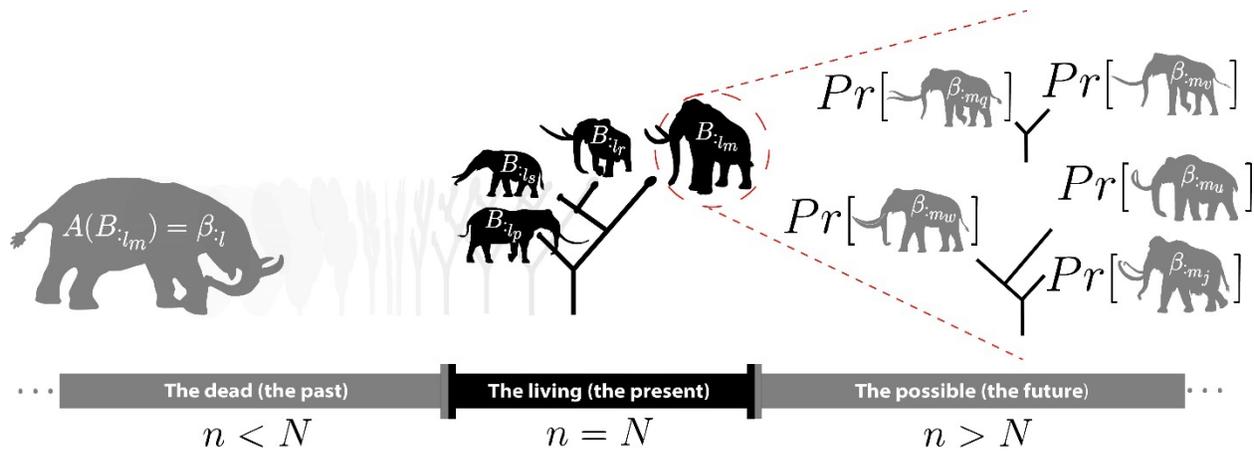

**Figure 3** *Paths through the tree of life.* Reproduction events are indexed by $n$, with the present denoted by $N$. Nested subscripts track lineage-specific ancestry.

A biological event $\beta_{:l}$ marks a point in lineage space; a random variable $B_{:l_m}$ tracks the evolving form of that lineage. Formally, each lineage is a stochastic process:

**Expression 1**

$$\left\{B_{:k_l}(n, \beta_{:k})\right\}, \ n \in \mathbb{N}, \ \beta_{:k} \in \mathbb{T}_n^{dead}.$$

For fixed $n$, $B_{:k_l}(n, \cdot)$ is a random variable over ancestral paths. For fixed $\beta_{:i}$, $B_{:k_l}(\cdot, \beta_{:i})$ gives a realization of biological form – either over the past ($\mathbb{T}_n^{\text{Dead}}$, Definition A1) or the future.

The nested structure of a lineage defines a unique **convex hull** in form space – the intersection of the supports of its ancestral probability distributions. This hull encodes evolutionary history and forms the basis for computing a path's entropy profile (Section 2.6.1):

**Equation 1**

$$conv(B_{:m}) = \bigcap_{n=1}^{N} [0, \beta_{:\alpha:}), \quad \alpha = 0, \ldots, A(B_{:m}).$$

Here, $\alpha$ indexes the ancestral path leading to $B_m$. Each interval $[0, \beta_\alpha)$ is the support of the ancestor's probability distribution – closed on the left to include death/extinction, and open on the right since no descendant is exactly identical to its ancestor. This preserves biological constraints while ensuring measurable boundaries.

## 2.2 Reproduction as recursively nested stochastic fission

Each reproduction event is a stochastic fission – a probabilistic branching in which a single entity gives rise to one or more descendants.

Formally, at each reproduction event $n$, the random variable $B_l$ generates a random number of descendants via a branching process. Simulation results (Section 3) use the Galton-Watson process (Equation B2). This branching process is used to model the dynamic tree of life as a **random iterated function system** (Appendix B), where the iterated function is the branching process itself. A random tree is generated at a scale between zero and one; for each leaf, a smaller-scale tree is recursively generated (visualized in Animations 1, 2). This yields a hierarchy of nested random fissions.

Each fission acts as a convex, generative transformation in form space, incrementally expanding the branching structure and shifting the topological dimension $D_T$ from 0 to 1 (Appendix A.2).

### 2.2.1: Assumptions of the model

The model relies on several simplifying assumptions. It excludes fusion (Appendix A.2), limiting the system to asexual reproduction; including fusion would require shifting from a strictly tree-based representation to a more general network structure, introducing transitions across multiple topological dimensions – points, lines, and planes. Each entity reproduces only once, at the end of its life (Appendix B), preventing overlap between ancestor and descendant generations; relaxing this constraint would require a formalism for partial convexity.

The implementation uses only uniform probability distributions (Expression 4; Section 2.6.1) with fixed limits on both generations and offspring (Appendix B). As a result, the model captures drift only, excluding selection and also mutation. This restriction is intentional: the aim is to isolate the structural features of evolution that emerge even without selective pressure, though it represents a simplified view of inheritance. All leaves within a given subtree are generated at the same scale (Appendix B). Relaxing this would add technical demands without substantially changing the biological interpretation (Appendix B.6).

Relaxing any of these assumptions would likely reinforce rather than undermine the central result: the multifractality of time. Greater lineage diversity and scaling variability would be expected to intensify, not diminish, the fractal structure. That multifractality emerges even under these minimal conditions suggests it is a robust and intrinsic feature of evolutionary dynamics. With the generative dynamics established, we can distinguish regions of the tree that are *alive* from those that are *dead*.

## 2.3 Life, death, and probabilistic change

In this model, life is defined by probabilistic change (Appendix A.3). An entity is **alive** if its future form is uncertain – i.e., if it is a random variable with nonzero probability of change (Definition A2). It is **dead** if its form is fixed with probability one (Definition A1). The collection of all such entities at the present reproduction $N$ forms the living tree of life, denoted $\mathbb{T}_n^{\text{Alive}}$. These are the $m$ with a well-defined nonzero dynamic $(\partial B/\partial t)_m$. This distinction reveals a deeper point: from within the system, the living not only persist – they define the present.

### 2.3.1 The living as the present

By Definition A2, the living are non-static: once realized, they probabilistically transition into new forms. This reflects both our simplifying assumption that entities reproduce only once, at the end of life (Section 2.2.1), and the actual structure of the tree of life – the living are always in flux.

If $N$ denotes the present, then no random variables with $n < N$ are living leaves. Once reproduction occurs again, the living tree becomes $\mathbb{T}_{N+1}^{\text{Alive}}$, and the previous $\mathbb{T}_N^{\text{Alive}}$ becomes the newest layer of $\mathbb{T}_N^{\text{Dead}}$. The living tree exists only in the present and consists entirely of entities in transition. Even so, any model of the living must rely on finite approximations.

### 2.3.2 The present as approximation

In biological reality, the present cannot be measured directly: once a lineage is observed, it has already shifted into the past. We capture this by treating the living as the large-reproduction limit of the dead.

Due to the convex probability structure, all chains of random variables converge in probability to zero as $n \to \infty$: death is inevitable. But at large finite $n$, these chains approximate the living entities at their dynamic endpoints. Life is therefore approximated in terms of what is already dead – a snapshot of the present that instantly becomes the past.

When $n$ is sufficiently large, this snapshot approximates the present well. We define the leaves – the living – as the large-reproduction limits over their unique paths:

**Equation 2**
$$B_{:i} = \lim_{n \gg 1} conv(B_{:i}).$$

Here we write $lim_{n \gg 1}$ rather than $lim_{n \to \infty}$ to emphasize that biological systems are inherently finite and never infinitely resolved. The approximation is therefore more than just a mathematical convenience – it reflects the biology itself. While the iterated function system has $\{0\}$ as its invariant set in the strict limit $n \to \infty$ (see Appendix D), lineages that are alive have not yet reached this terminal state. Instead, they are approximately converged for large but finite $n$, a notion made precise in Lemma D.2. The approximate limit is therefore both mathematically valid and biologically meaningful: to be alive is to be not yet ended.

**The living tree of life** is thus the large reproduction limit over the dead tree of life:

$$\mathbb{T}^{\text{Alive}} = \lim_{n \gg 1} \mathbb{T}^{Dead}_n.$$

Equation 3

Appendix D gives the details of this convergence. These finite-but-large limits define the living as paths still in motion – stochastic processes that remain probabilistically open.

### 2.3.3 Observers as the living

In the large reproduction limit (Appendix D.2), each lineage culminates in a living entity – any $m$ for which the dynamic $(\partial B / \partial t)_m$ is well-defined and nonzero (Definition A2). We refer to such entities as **observers**. Each observer has a probability density function $f_m$ (Section 2.6.1) determined by the convexity of its unique ancestry, giving every living path through the tree a distinctive informational fingerprint in phylogenetic space.

## 2.4 The tree of life is multifractal

Having defined the tree of life as a dynamic, probabilistically nested structure built through recursive fissions, we now examine its geometric consequences. The resulting architecture is multifractal (Appendix C): each lineage traces a distinct monofractal curve shaped by its unique ancestral path, while the full tree forms an ensemble of such curves (Hudnall & D'Souza, 2025). In this section, we summarize the quantitative framework for analyzing this structure using fractal dimensions.

### 2.4.1 Fractal dimensions

We capture the internal scaling geometry of each lineage using a **fractal dimension**. Specifically, we apply the correlation dimension (Grassberger & Procaccia, 1983), which estimates dimensionality from the distribution of pairwise distances (Appendix E). The fractal dimension of path a $B_i$ is denoted: $D_F(B_i)$.

To compare time between lineages we must also identify the fractal dimension at which their paths diverged – i.e., the resolution at which they split from a shared ancestor.

### 2.4.2 The crosspath fractal dimension

For any pair of entities, we define the **crosspath fractal dimension** as the dimension of their most recent common ancestor:

Equation 4

$$D_F\left(B_{:i}, B_{:j}\right) = D_F\left(A\left(B_{:i}, B_{:j}\right)\right).$$

This value marks the scale at which the lineages first become distinguishable and their entropy profiles begin to diverge. It serves as the structural reference point for computing lineage-relative time intervals and as a key geometric input to the dilation equation in Section 2.7. In this way, the multifractal geometry established here directly informs the temporal comparisons that follow. To ensure validity, we apply a convergence criterion to exclude poorly resolved or unstable paths from consideration.

## 2.5 Convergence tolerance

The correlation dimension is meaningful only for paths of sufficient length. At early generations, the estimate is unstable or undefined. To ensure reliability, we restrict analysis to paths with reproduction depth $n$ large enough that the estimated dimension has converged toward a limiting value.

In simulation, this **convergence tolerance** is enforced by accepting only those paths for which the estimated fractal dimension differs from its final value by less than 10%:

**Expression 2**

$$|D_F(\text{n}) - D_F(\text{final})| < 0.1 \times D_F(\text{final}).$$

Appendix D.3 provides a sensitivity analysis around this 10% threshold, showing that it is conservative and ensures high-resolution approximation, selecting only well-converged paths.

### 2.5.1 Convergence tolerance for crosspath fractal dimensions

When computing $D_F(B_i, B_j)$, the estimate is only accepted if both lineages satisfy the convergence threshold at their shared ancestor. This strict filter greatly reduces the number of valid comparisons (Section 3.2.2), reflecting a fundamental feature of the tree's multifractal geometry: the deeper the divergence, the less resolved the structure becomes.

Having established the mathematical foundation for lineage-level scaling, we now turn to the information-theoretic framework used to quantify multifractal time intervals.

## 2.6 Information measures

To define observer-relative time within the multifractal tree, we use standard information-theoretic measures (Cover & Thomas, 2006) to quantify uncertainty, constraint, and similarity across lineages. These quantities provide a rigorous mathematical basis for comparing random variables and have been widely applied in biological and stochastic systems (Adami, 2004; Chanda et al., 2020; Contreras-Reyes & Kharazmi, 2023). We begin by formalizing entropy along paths, then introduce mutual information and information distance to compare lineages.

### 2.6.1 Entropy of biological form

As reproduction proceeds, the set of possible forms becomes increasingly constrained (Appendix A.1). This reduction in uncertainty can be quantified using entropy, which captures the lineage-specific progression of form constraint.

*The probability density function*

Each lineage is modeled as a path of nested random variables with shrinking supports (Appendix A.3). In the drift case (Section 2.2.1), where fitness differences are absent, each random variable is uniformly distributed over the support defined by its immediate ancestor:

**Expression 3**

$$B_{0:l_m} \sim U\left(0, A\left(B_{0:l_m}\right)\right), \text{ with } Pr(B_0 = 1 = \beta_0) = 1.$$

The initial condition sets the ancestor of all life $\beta_0$ to 1 with certainty. The corresponding **probability density function** $f_m$ is:

**Equation 5**

$$f_m(x) = \frac{1}{A\left(B_{:l_m}\right)} = \frac{1}{\beta_{:l}}, \qquad x \in \left[0, A\left(B_{:l_m}\right)\right).$$

This density function forms the basis for calculating path entropies.

*Path entropies*

For any path $\beta_{:l}$ leading to the random variable $B_{:l_m}$, the differential **entropy** is:

**Equation 6**

$$H[B_{:m}] = -\int_0^{A(B_{:m})} f_m(x) log(f_m(x)) dx,$$

Which under Equation 5 simplifies to:

**Equation 7**

$$H[B_{:m}] = log\left(A(B_{:m})\right) \leq 0.$$

Entropy is strictly nonpositive and becomes increasingly negative with each generation, reflecting accumulated constraint. This matches long held expectations that living systems are negative entropy producing structures (Schrodinger, 1944). Entropy is undefined for the first entity (no ancestor) and zero for the second (since $\beta_0 = 1$), mirroring the no-resolution regime of fractal dimension (Section 2.5).

### 2.6.2 Crosspath quantities

To compute relative time intervals between lineages, we extend beyond the pathwise entropies and introduce **crosspath quantities**: symmetric measures that capture divergence between lineages descending from a shared ancestor but lacking a direct ancestral link.

The three measures central to our framework are the **joint entropy** $H[B_i, B_j]$, which quantifies the total uncertainty across both lineages; the **mutual information** $I[B_i; B_j]$, which captures the shared uncertainty between them; and the **information distance** $d_\beta(B_i, B_j)$, which gives a normalized measure of separation in information space. **Figure 4**, contrasts pathwise and crosspath quantities.

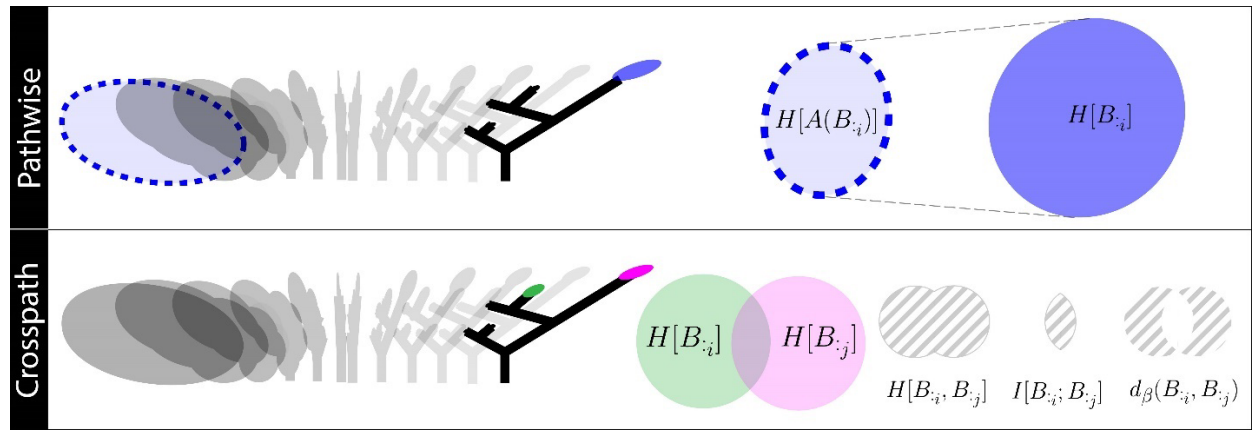

**Figure 4** *Pathwise versus crosspath quantities.* (**TOP**) The path entropy $H[\cdot]$ tracks single lineages. (**BOTTOM**) Crosspath quantities track informational relations between pairs that do not share the ancestor-descendant relationship.

*The joint probability density function*

If $B_i$ and $B_j$ diverge from ancestor $A(B_i, B_j)$, with supports $[0, A(B_i))$ and $[0, A(B_j))$ respectively, then their joint probability is the product of their conditional probabilities given their ancestor:

Equation 8
$$Pr(B_i \leq x, B_j \leq y) = Pr(B_i \leq x \mid A(B_i, B_j) = z \cap B_j \leq y \mid A(B_i, B_j) = z).$$

Here, both terms inside the conditional are events, joined by intersection. Since $B_i$ and $B_j$ are conditionally independent given their common ancestor, their **joint distribution function** in the drift case is:

Equation 9
$$F(B_i, B_j) = \frac{xy}{A(B_i, B_j)^2}, \quad x \in [0, A(B_i)), \quad y \in [0, A(B_j)).$$

And so their **joint density function** is then:

Equation 10
$$f_{ij}(x, y) = \frac{1}{A(B_{\cdot i}, B_{\cdot j})^2}.$$

*Joint entropy*

The **joint entropy** for any two leaves $B_i$ and $B_j$ is given by:

Equation 11
$$H[B_{\cdot i}, B_{\cdot j}] = -\int_0^{A(B_{\cdot i})} \int_0^{A(B_{\cdot j})} f_{ij}(x, y) \log\left(f_{ij}(x, y)\right) dxdy.$$

Integrating over each leaf's support, this simplifies to:

Equation 12
$$H[B_{\cdot i}, B_{\cdot j}] = \frac{2A(B_{\cdot i})A(B_{\cdot j})}{A(B_{\cdot i}, B_{\cdot j})^2} \cdot \log\left(A(B_{\cdot i}, B_{\cdot j})\right).$$

As with single-lineage entropy, the joint entropy is a strictly nonpositive quantity, reflecting the use of continuous random variables over bounded sub-intervals of $[0.1]$. The tighter the shared support, the more negative the entropy becomes — indicating lower uncertainty and finer ancestral resolution.

*Mutual information*

The shared dependence of any two leaves $B_i$ and $B_j$ is given by their **mutual information** $I$:

Equation 13
$$I[B_{\cdot i}; B_{\cdot j}] = \int_0^{A(B_{\cdot i})} \int_0^{A(B_{\cdot j})} f_{ij}(x, y) \log\left(\frac{f_{ij}(x, y)}{f_i(x) f_j(y)}\right) dxdy.$$

Substituting Equation 10, this evaluates to the nonpositive quantity:

Equation 14

$$I\left[B_{:i}; B_{:j}\right] = \frac{A(B_{:i})A\left(B_{:j}\right)}{A\left(B_{:i}, B_{:j}\right)^2} \log\left(\frac{A(B_{:i})A\left(B_{:j}\right)}{A\left(B_{:i}, B_{:j}\right)^2}\right).$$

*Information distance*

For any two biological forms $B_i$ and $B_j$, their informational separation is given by their **information distance** $d_\beta$ is given by:

Equation 15

$$d_\beta\left(B_{:i}, B_{:j}\right) = H\left[B_{:i}, B_{:j}\right] - I\left[B_{:i}; B_{:j}\right].$$

This value captures the total uncertainty that is unique to each lineage beyond what is shared. While similar in form to the variation of information, we avoid that term here because our variables are continuous. For discrete random variables, $d_\beta$ satisfies the properties of a true metric (Cover & Thomas, 2006). For continuous random variables, as in this work, strict metricity does not hold: $d_\beta$ can take both positive and negative values, making it a bidirectional measure of lineage-specific uncertainty rather than a true metric. (see Section 3.3.3).

## 2.7 Making observer-relative time measurable

We now derive a dimensionless equation for multifractal time intervals using information-theoretic quantities, thereby operationalizing the claim that time is a function of form. For clarity, explicit references to random variables are dropped, using indices alone (e.g., $i$ instead of $B_i$).

### 2.7.1 Nottale's scale dilation equation

To compare the relative time elapse between two lineages, we evaluate the lengths of their paths back to their most recent common ancestor. For this, we adapt Nottale's scale dilation equation (Nottale, 1993), a formalism for measuring relative distances in scale-dependent geometries. Reinterpreted here, it yields a dimensionless ratio of time elapse between two observers based on their positions in the tree.

Let $i$ and $j$ be any two leaves in the system. Their relative time elapse is the ratio of their path lengths from the present back to their common ancestor (shown in Figure 2):

Equation 16

$$\frac{\Delta t_i}{\Delta t_j} \equiv \partial t_j^i.$$

Assuming higher entropy corresponds to greater time elapse, if $|H[i]| > |H[j]|$ then:

Equation 17

$$\partial t_j^i = 1 + \Gamma,$$

where $\Gamma > 0$ and depends on the relative locations of $i$ and $j$ in the tree. If instead, $|H[i]| < |H[j]|$, then the inverse is taken.

In Nottale's formulation, the dilation factor $\Gamma = (\lambda/\varepsilon)^{D_F - D_T}$, where $\varepsilon$ and $\lambda$ are respectively the observation and transition scales, $D_F$ is the fractal dimension, and $D_T$ is the topological dimension. Next, we reinterpret these terms through the lens of information theory.

### 2.7.2 The information-theoretic dilation equation

In biological systems, scale is determined not by raw geometry but by inherited structure along ancestral paths—a relationship captured by information-theoretic quantities. We replace the geometric terms in the dilation equation with informational analogs: the transition scale $\lambda$ becomes the mutual information $I[i;j]$ between two lineages, reflecting their most recent common ancestor; and the observation scale $\varepsilon$ becomes their information distance $d_\beta(i,j)$, quantifying their divergence.

Substituting these quantities into Equation 17, along with our fractal formalism (Section 2.4.2), yields an **information-theoretic dilation equation** – an expression of the relative time elapse between observers:

**Equation 18**

$$\partial t_j^i = 1 + \left(\frac{I[i;j]}{d_\beta(i,j)}\right)^{D_F(i,j)-1}, \quad |H[i]| > |H[j]|, \quad i \nsim j$$

Here, $D_F(i,j)$ is the crosspath fractal dimension (Equation 4), and $D_T = 1$ for trees (Appendix A.2). The condition $i \nsim j$ denotes that $i$ and $j$ are not in an ancestor-descendant relationship: their convex hulls are not nested – i.e., the ancestral supports used to construct $conv(i)$ and $conv(j)$ do not contain one another. If $|H[j]| > |H[i]|$, then we take the inverse. If $|H[i]| = |H[j]|$ then there is no criterion to choose a direction. This occurs when $i$ and $j$ share an ancestor within two generations (Appendix A.3.1). In these cases $I[i;j] \to I[i;i] = H[i]$, $d_\beta(i,j) \to d_\beta(i,i) = 0$, and $D_F(i,j) \to D_F(i,i) = D_F(A(i))$, yielding:

$$\partial t_j^i \approx 1 + \left(\frac{0}{H[i]}\right)^{1-D_F(A(i))} = 1.$$

This indicates approximate agreement in time elapse. Hence, Equation 18 behaves as expected: for distant lineages (large $d_\beta$, small $I$), $\partial t_j^i \gg 1$; for closely related lineages (large $I$, small $d_\beta$), $\partial t_j^i \approx 1$.

This dilation equation provides the mathematical foundation for a general theory of measurement grounded in biological lineages. Before applying it, we require that the crosspath fractal dimension satisfy the convergence criterion defined in Section 2.5.1: both lineages must have individually converged. This ensures that the scale $D_F(B_i, B_j)$ anchoring the calculation is well resolved and stable, preventing poorly resolved or unstable paths from introducing spurious results.

### 2.7.3 The living tree as a nearness space

In its multifractal geometry, branchwise "distances" are path- and scale-dependent: without fixing an observer, the same pair of entities can yield different separation values, violating well-definedness and the metric axioms. The intrinsic geometry is instead captured by a *nearness relation* given by the dilation equation: $i$ and $j$ are near if their $\partial t_j^i$ value is nonzero, with real values indicating divergent histories and complex values indicating coherent histories (Section 3.3.3). Magnitudes and arguments of $\partial t_j^i$ grade this nearness without assigning absolute lengths (Section 3.4).

Equation 18 thus defines these $\partial t_j^i$ values as dimensionless ratios based entirely on lineage information and fractality. Viewed externally, the tree of life is a *nearness space* rather than a metric space: proximity and continuity are well-defined, but absolute temporal distances are not. To recover measurable durations, the system must be centered on a specific observer.

### 2.7.4 The emergence of local measures: centering the tree

**Centering** anchors the system to an observer $p$, converting nearness relations into measurable durations. Restricting to all pairs involving $p$ ensures $\partial t_p^p = 1$, with all other intervals expressed as fractions or multiples of this unit:

**Expression 4**

$$\{\text{All } \partial t_j^i \mid i = p \text{ or } j = p\}.$$

This defines the operational unit of time for $p$ and makes cross-species time measurement possible in principle. For it to be actual, the centered observer must also be able to define operational time units – such as humans with the SI second (Section 3.5.2). Full formalism and proofs of the nearness-space properties are given in $\text{Appendix F}$.

### 2.7.5 Clarification: distinction from non-ultrametric phylogenies

It is important to distinguish the present framework from the standard phylogenetic claim that empirical trees are non-ultrametric. In phylogenetics, an *ultrametric tree* is one in which all tips are equidistant from the root, corresponding to a strict molecular clock. When rates vary among lineages, root-to-tip distances differ and the tree is classified as *non-ultrametric*. Crucially, both ultrametric and non-ultrametric trees still presuppose a global metric: time is treated as a universal background in which rates can be faster or slower along different branches.

Our construction departs more fundamentally. We do not assume a global metric at all. The recursive branching process generates a multifractal geometry that fails the metric axioms unless centered on a specific observer. Externally, the tree is not simply "non-ultrametric"; it is non-metric. Internally, observer-relative centering induces a local measure, yielding dilation ratios that recover measurable durations in standard units (Section 3.5). In this sense, our claim is categorical rather than parametric: rather than adjusting rates within a background metric, we show that the metric itself does not exist globally and must be constructed from within.

# 3 Results and discussion

This section simulates the framework developed in Section 2, synthesizing its outputs into a cohesive theory of time as an emergent, measurable quantity rooted in phylogenetic structure. Simulations show that nested, stochastic reproduction produces a multifractal architecture in which each lineage traces a unique trajectory shaped by its informational ancestry. These paths exhibit distinct fractal dimensions, entropy profiles, and informational relationships, jointly expressing the asymmetries and divergences predicted by the theory. The information-theoretic dilation equation (Equation 18) quantifies observer-relative time intervals, confirming that time within the living tree is not universal but depends on the lineage-specific accumulation of information (Section 3.4).

## 3.1 Approximating the living: the dynamic tree in simulation

Each lineage emerges as a nested sequence of dependent random variables produced through recursive fissions (Sections 2.1, 2.2). These fissions create convex topological transformations in form space, inducing lineage-specific dynamics $\partial B_m / \partial t_m \equiv (\partial B / \partial t)_m$. Observers are those entities $m$ for which this dynamic is nonzero – the large-reproduction limits along their ancestral paths (Section 2.3).

**Figure 5** shows observers in simulation. A total of 976,230 paths were generated through 30 iterations of the random iterated function system, with terminal nodes approximating observers (Equations 2, 3). From these, a random sample of 10,000 paths is displayed.

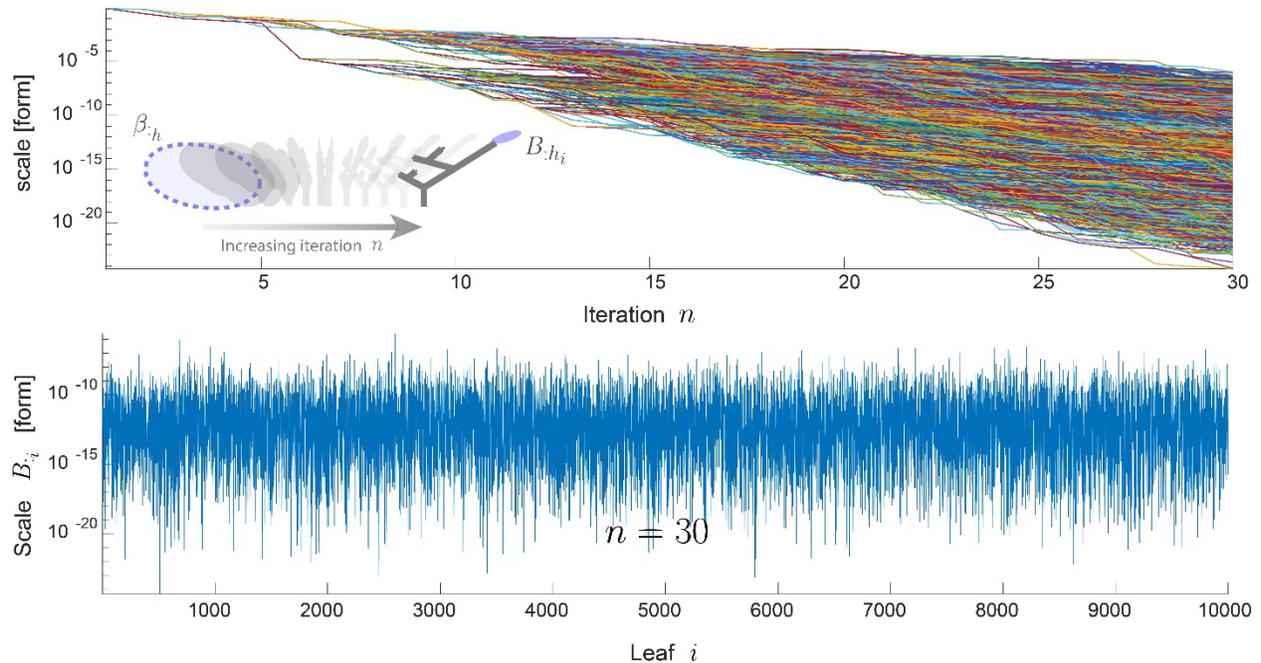

**Figure 5** *Paths and observers in the tree of life.* (**TOP**) Biological form variables $B_{:i}$, are mapped to the $n$-dimensional unit interval as nested sequences of random scale variables. (**BOTTOM**) Observers (the living) are approximated as terminal reproductions after 30 iterations.

This simulation shows the living tree as an ensemble of converged stochastic branching paths. With observers established, we now turn to their central structural consequence: the tree of life is multifractal.

## 3.2 Verified multifractality in simulation

Multifractality arises directly from the principles of nestedness, duality, and randomness, and accounts for the observer-relativity of time – the duration of experience varies across entities according to their informational structure (Hudnall & D'Souza, 2025). This section verifies multifractality in simulation.

### 3.2.1 Pathwise fractal dimensions

While multifractality can be assessed via generalized Hurst indices (Appendix E.1), the spectrum of lineage-dependent fractal dimensions (Section 2.4) offers a more direct and interpretable measure. **Figure 6 TOP** shows the pathwise fractal dimensions, revealing wide variation in scaling behavior.

### 3.2.2 Crosspath fractal dimensions

The crosspath fractal dimensions (Equation 4) were computed for all leaf pairs meeting the convergence requirements of Section 2.5.1. In our simulated system, about 0.001% of the nearly 477 billion possible comparisons met this threshold. These results are statistically robust ($\text{Appendix } G.1$). **Figure 6 BOTTOM** shows their distribution, confirming that both individual lineages and shared evolutionary histories exhibit non-uniform scaling.

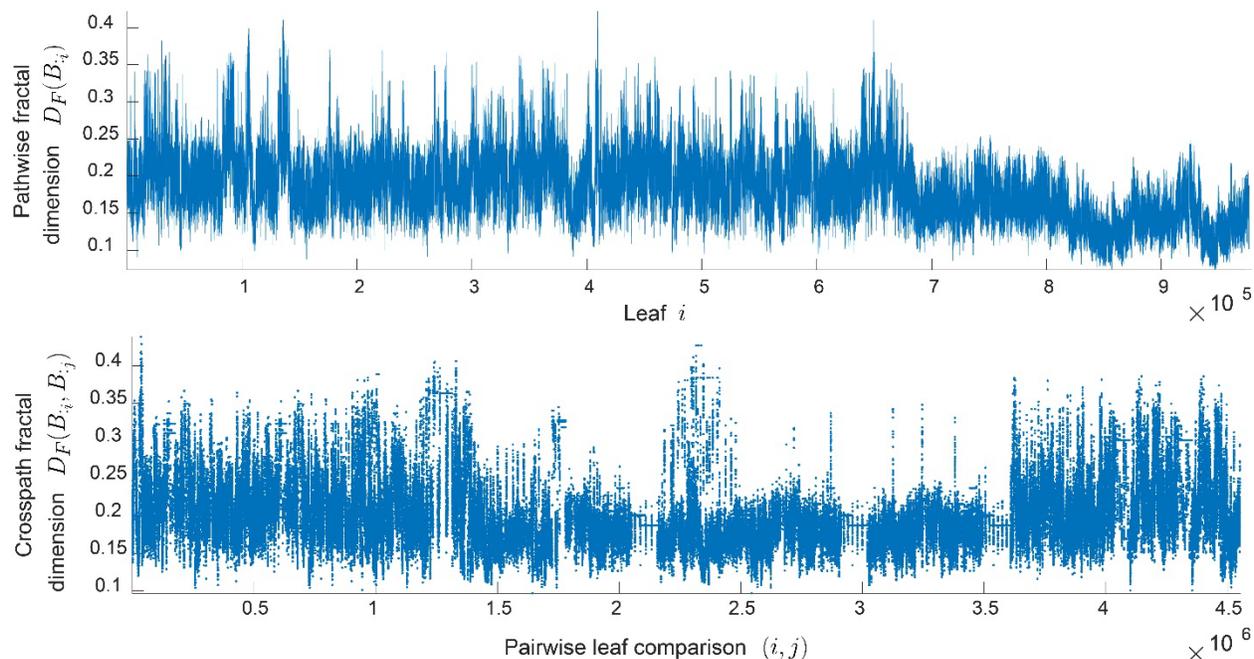

**Figure 6** *Fractal dimensions.* (**TOP**) Pathwise fractal dimensions for each of the 976,230 simulated lineages from Figure 5. (**BOTTOM**) Crosspath fractal dimensions for all convergent (Section 2.5.1) pairwise leaf comparisons.

Despite their relative sparsity, the strict convergence filter ensures these crosspath dimensions are well-resolved. We now turn to the information-theoretic results.

### 3.3 Information quantities

To make observer-relative time measurable, we must first quantify the informational structure of the tree. Each lineage carries a signature of its evolutionary history in the form of path entropy (Section 2.6.1), while crosspath comparisons use joint entropy, mutual information, and information distance (Section 2.6.2). These quantities formalize how form becomes constrained through descent, how lineages relate to one another, and how coherent or divergent their evolutionary paths have been. Together, they constitute the backbone of the time dilation equation (Section 2.7.2).

### 3.3.1 Path entropies

Entropy trajectories for individual lineages (Equation 7) track how reproduction progressively constrains future possibilities. **Figure 7** shows this accumulation for the same 10,000 randomly sampled paths from Figure 5, both as full trajectories and as final entropy values at the living leaves (observers).

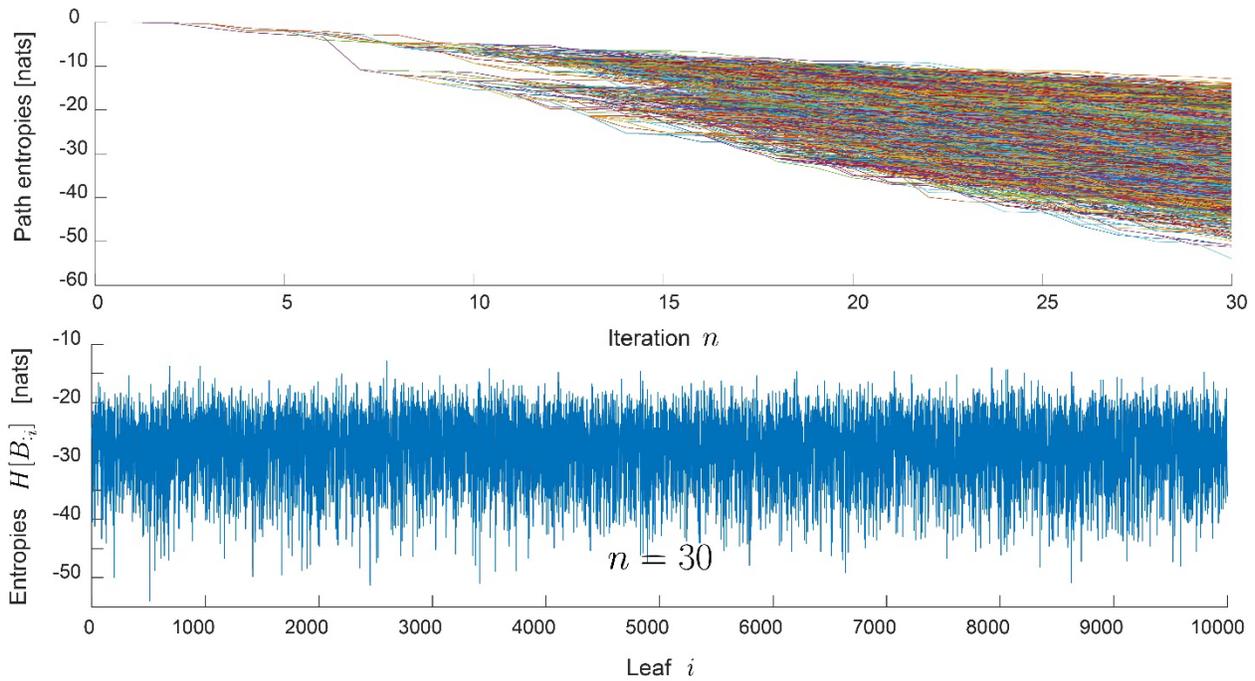

**Figure 7** *Path entropies. (TOP)* Entropy of all 10,000 paths from $n = 1$ to 30. (**BOTTOM**) Entropy at the observers ($n = 30$).

This directional asymmetry – entropy becoming increasingly negative over reproductions – captures a fundamental signature of lineage evolution. whereas thermodynamic entropy increases with disorder, biological entropy increases in magnitude in the negative direction: lineages become more ordered and specific as each reproduction step narrows the set of possible future forms.

### 3.3.2 Joint entropy and mutual information

Joint entropy (Equation 12) measures combined uncertainty of two lineages; mutual information (Equation 14) measures how much of that uncertainty is shared. Computed for all leaf pairs at $n = 30$, these quantities characterize the coherence or divergence between lineages. **Figure 8** shows results for the convergent pairs (Section 2.5.2) from the nearly 477 billion possible comparisons.

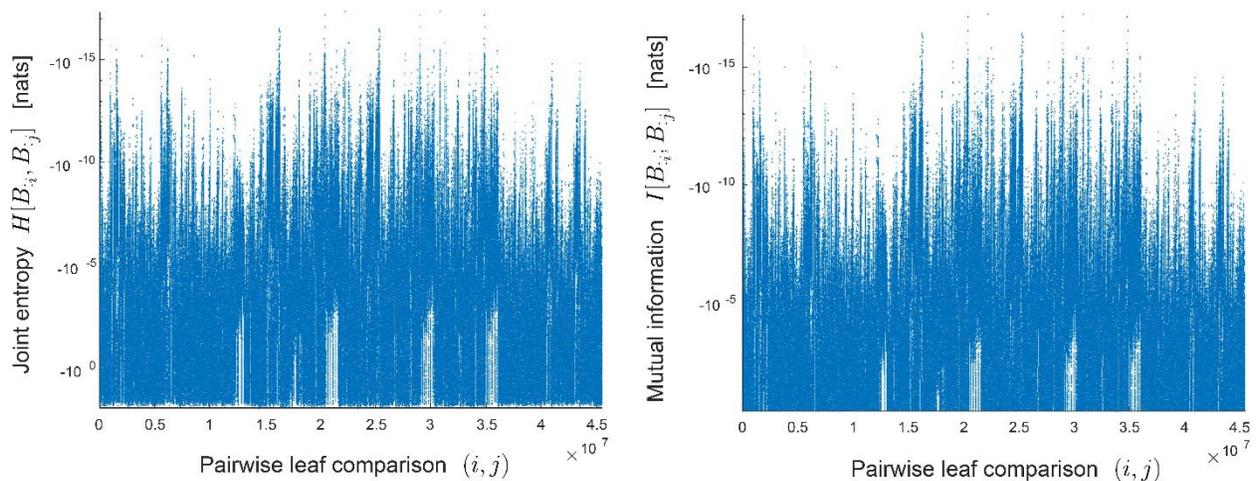

**Figure 8** *Joint entropy and mutual information.* (**LEFT**) Joint entropy and (**RIGHT**) mutual information for convergent pairs.

Because mutual information is strictly nonpositive, large magnitudes indicate strong evolutionary similarity between lineages, while values closer to zero reflect greater divergence. Although the distributions in Figure 8 appear similar, their distinction becomes clear via the information distance.

### 3.3.3 Information distance

The information distance $d_\beta$ integrates joint entropy and mutual information into a single measure of phylogenetic separation (Equation 15).

*Coherent versus divergent evolution*

For discrete variables, $d_\beta$ is nonnegative and metric (Cover & Thomas, 2006). In our model, convex mappings of continuous random variables over contracting supports (Appendix A.3.1) allow both positive and negative values, with the sign carrying meaning: positive values indicate informational coherence, negative values indicate divergence. **Figure 9** presents the distribution of $d_\beta$.

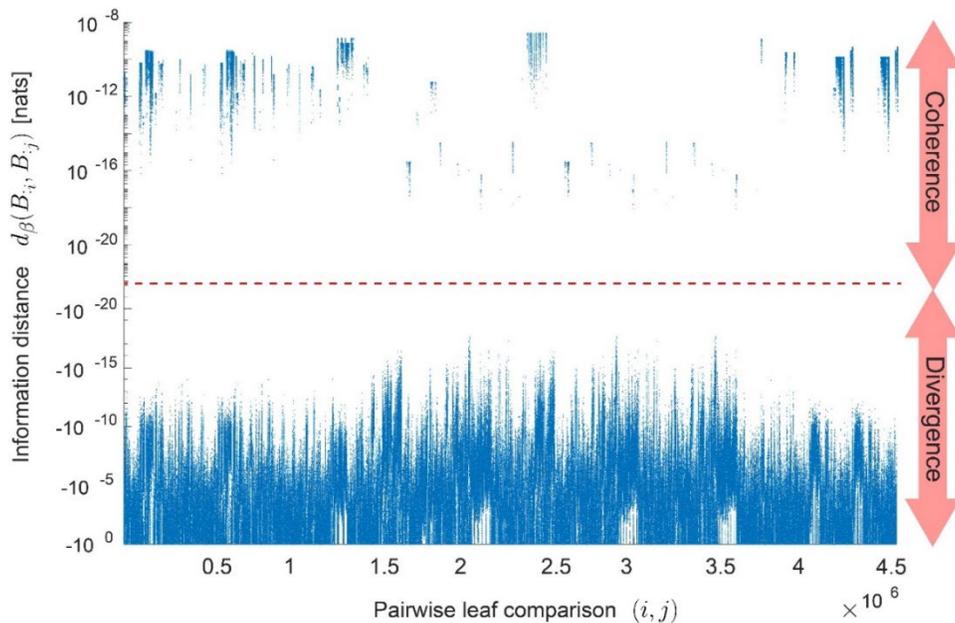

**Figure 9** *Information distances between observers.* Distribution of $d_\beta$ for the convergent leaf pairs of Figure 8. Positive values (99.4% of points) indicate coherence – mutual information dominates, meaning lineages remain aligned in their constraints. Negative values (0.6%) reflect divergence, where joint entropy dominates and ancestral constraints have parted. This skew reflects the baseline expectation under neutral drift that most lineages remain similar in the absence of selection (Section 2.2.1). The red dashed line marks the approximate transition; note the asymmetric log scales. (These results are statistically robust – see Appendix G.2.)

We now integrate the information-theoretic and fractal dimension results to compute relative time intervals. Equation 18 then yields a dimensionless, observer-relative time interval based entirely on lineage information and fractality – a nearness ratio that converts between the temporal experiences of different observers.

### 3.4 Simulation results for the dilation equation

Having established the information structure of the tree, we now use these quantities to calculate time dilation between lineages.

The sign of $d_\beta$ determines whether observer-relative time intervals (Equation 18) are real or complex. When $d_\beta(i,j) < 0$, indicating divergent evolution, the resulting dilation values are real and yield temporal conversions factors between lineages. When $d_\beta(i,j) > 0$, indicating coherent evolution, the dilation values are complex, appearing as points in the complex plane. **Figure 10** shows the convergent solutions derived from the system in Figure 5.

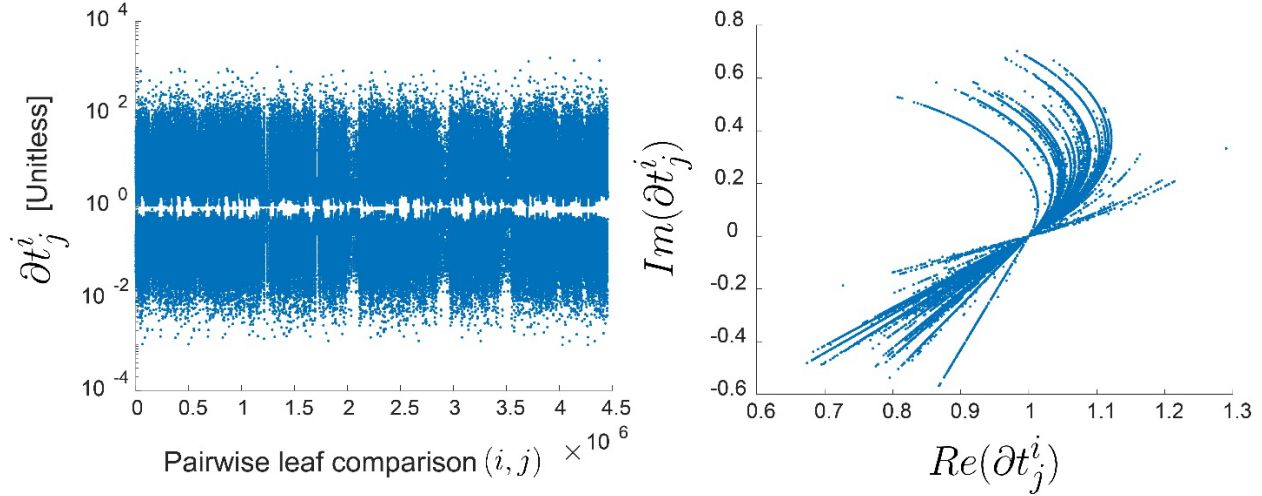

**Figure 10** *Solutions to the dilation equation.* Of the 477 billion pairwise comparisons, 4,551,026 passed the convergence threshold. (**LEFT**) 4,459,514 real-valued solutions (∼ 99.4%) correspond to divergent lineages. (**RIGHT**) 91,512 complex-valued solutions (∼ 0.6%) correspond to coherent lineages. Points cluster around the attractor at $(Re, Im) = (1,0)$, where $\partial t_j^i = 1$, marking agreement in time elapse. Solutions in the upper and lower half-planes are inverses: if $\partial t_j^i$ lies in $Im > 0$, then $\partial t_i^j$ lies in $Im < 0$. Lower-half points align by phase angle (radial lines), while upper-half points align by magnitude (circular arcs).

These distributions are statistically robust (Appendix G.4.1). While strict convergence filtering eliminates most comparisons, real solutions are more likely to satisfy the criteria: fewer than 0.0001% pass the filter in the coherent case, compared to ∼ 0.14% in the divergent case (Appendix G.4.2).

As shown in Figure 10, complex-valued dilation ratios appear as coherent structures in the attractor. While the present analysis proceeds with real-valued solutions, the complex solutions indicate additional structure that will be taken up in future work (see Section 3.6.1).

With the dilation equation evaluated, we now shift from global structure to the *measurement* of time.

### 3.5 Measuring time

Throughout this work, "observer" has referred to anything alive – any lineage in the tree with a nonzero dynamic (Section 2.3). To define time operationally, we now introduce a **measurer**: an entity that not only observes but defines units, making explicit comparison possible. The key insight is that, due to multifractality, any defined unit is necessarily specific to the form of its definer. This section shows how observer-relative quantities become measurable when anchored to such a measurer.

#### 3.5.1 Centering the tree to a biological observer

Centering (Section 2.7.4) the solutions to Equation 18 means setting $\partial t_i^i = 1$ for a particular $i$ in the system. In principle, the system can be centered on any member entity. In practice, one only gets to center the system on oneself. This is not merely philosophical: humans cannot experience the world as dogs do. But it also means that measurement becomes possible from a localized position in the tree.

Before centering, we apply the same stability filter described in Section 2.5: only those lineage pairs that meet the convergence tolerance, are retained. This ensures that the observer's "anchoring" is applied only to valid, converged solutions of the dilation equation.

### 3.5.2 From observation to measurement

Centering is necessary for measurement, but not sufficient. For measurement, the centered observer must also be a **measurer** – an entity capable of defining a unit, like humans and the SI second.

Solutions to the dilation equation (Equation 18) yield relative time elapse between lineages. But to compare durations absolutely, one must define a unit of time – a choice that implicitly selects a reference position in the tree. Thus, time measurement becomes actual only when the centered observer defines a unit.

When that condition is satisfied, as it is for us in the SI system, then at 9,192,631,770 Hz we get the definitional relation:

**Definition 1**

$$\partial t_{human}^{human} \equiv \Delta t_{human} \equiv 1 \text{ second.}$$

This anchors the tree of life as a **measure space** centered on the human lineage (Appendix F.3). The other dilation ratios $\partial t_{human}^{i}$ then give some fraction or multiple of the second, interpreted as the **rate of time elapse** for entity $i$ relative to a human.

**Figure 11** colloquially illustrates the biological dependence of measurement by centering on the human. The human defines a unit of time – the SI second using a cesium clock. This sets $\Delta t_{human} = 1$ second. The dilation ratio $\partial t_{human}^{dog}$ then gives the dog's rate of time elapse relative to the human. In this example, $\partial t_{human}^{dog} > 1$, meaning time elapses faster for a dog than for a human.

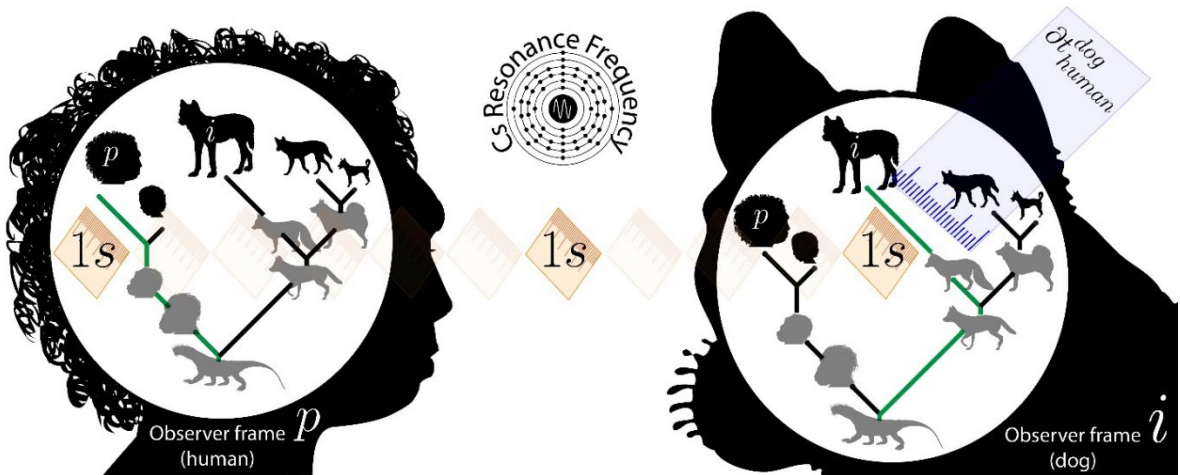

**Figure 11** *Measurement is biologically specific.* Here, the human measurer $p$ defines 1 *second* via the SI standard. Due to multifractality, the duration experienced within this human-defined unit is not uniform across the tree of life for all observers. The dilation ratio $\partial t_{human}^{dog}$ is a dimensionless conversion factor giving how fast time elapses for a dog relative to a human.

Figure 11 makes explicit what temporal theory has long left implicit: time units are biologically defined. Our results suggest that unit intervals do not propagate uniformly across life, but instead transform as a function of the observer's lineage. Although the figure is schematic, its logic applies directly to the SI second. The measurer is unavoidably human and cannot be abstracted out of the definition. The definition must make *explicit* reference to the human as the definer.

*But the SI second is nonetheless adequate*

Once a unit is defined, the dilation ratios (Equation 18) act as conversion factors, enabling cross-lineage comparisons. For example, if a human measures an elapsed time $\Delta t_{human}$ in seconds, then a dog will observe the quantity:

**Expression 5**

$$\Delta t_{human} \, \partial t^{dog}_{human},$$

also in seconds. The unit is fixed; the duration of time within it depends on lineage.

Because this formalism is dimensionally consistent, with $\partial t^i_j$ dimensionless, all units of standard science are preserved. Thus, the second is still operational — not as a universal quantity, but as a human-specific reference scale. Constants like the cesium resonance frequency remain invariant, but the duration of the unit interval varies across biological form. Established science is not violated, only *extended*.

### 3.5.3 Relative time measurements in simulation

In simulation, we perform centering by randomly selecting an observer $p$ and restricting the set of dilation ratios to those involving $p$ directly (Expression 5). This is more than a mathematical simplification; it reflects an empirical truth: observers have access only to their own experiences.

If $p$ is taken to be human, each centered dilation value $\partial t^i_p$ gives the rate at which time elapses for some entity $i$ relative to the human. **Figure 12** presents these centered values for the system of 976,230 leaves from Figure 5, with $p$ selected at random and conceptually treated as human. The result is a depiction of how time flows across the tree from within, relative to a single, embedded observer.

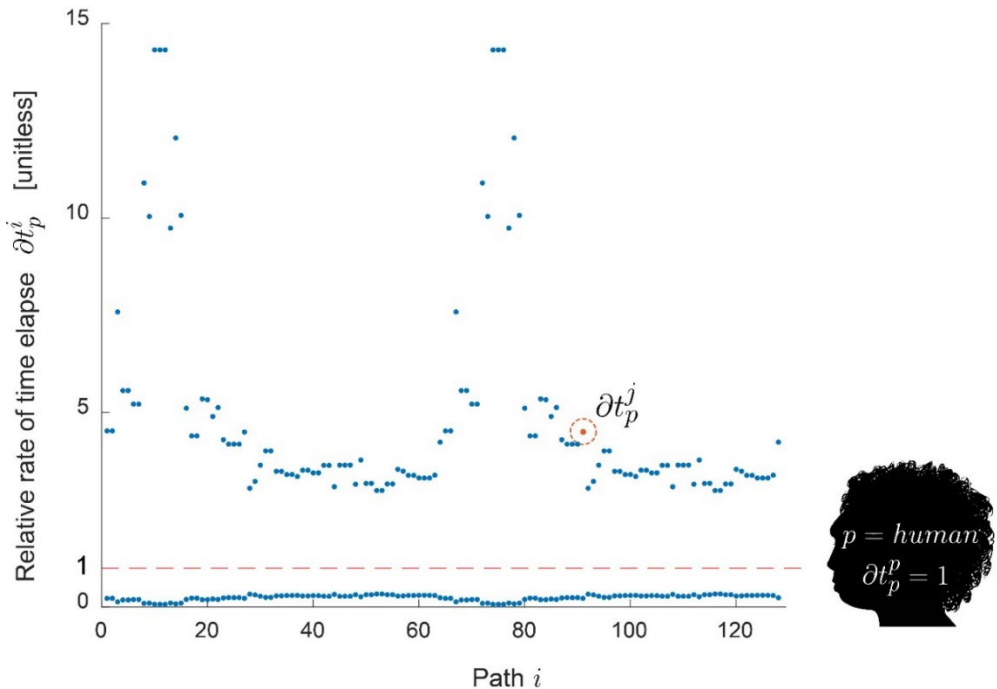

**Figure 12** *Relative rates of time elapse.* The dilation values $\partial t_p^i$ for 128 entities relative to a centered observer $p$ (conceptually human). Highlighted entity $j$ experiences time roughly 5 times faster than $p$.

Of the 976,230 total paths, valid dilation values could be computed for 128 ($\sim$ 0.013%). This small number reflects the compounding effects of three filters. First, convergence constraints on fractal dimensions (Section 2.5), eliminate most comparisons. Second, about 0.6% of the valid solutions are complex-valued, corresponding to coherent lineages that have not diverged in information space (Section 3.4). Third, a small subset of comparisons involve lineages whose most recent common ancestor is within two generations, yielding dilation values close to one (Section 2.7.2; statistics in Appendix G.3).

The centering procedure also reveals a strong heterogeneity in observational capacity across leaves. As shown in Appendix G.5, the number of observable target leaves varies over several orders of magnitude, with most leaves able to observe only a small fraction of the system and a few leaves capable of observing a disproportionately large number of targets. This distribution reflects the asymmetric information geometry of the tree.

Despite their small number, the centered values in Figure 12 render the model concrete: they give observer-relative time elapse as a measurable quantity derived from phylogenetic structure.

### 3.5.5 Time as a relational, measurable, and biologically anchored quantity

These values complete our construction of a biologically grounded, information-theoretic model of time and measurement. Rather than assuming a universal temporal metric, we have shown how time elapse can be quantified from lineage structure alone. By centering the model on a single observer, we recover the empirical condition of life: time as experienced from within the living system.

The dilation equation thus reconciles the relativity of time in biology with the operational need for fixed units. The second – though human-defined – becomes extensible across life through dimensionless ratios, preserving both empirical rigor and biological specificity. This framework makes time measurable from lineage structure alone.

The next section gives a preliminary look at the empirical tractability of this framework – not only as a theory of observer-relative time, but as the basis for a new scientific field: biokinematics.

## 3.6 Outlook: future directions

This expansion from theoretical construction to empirical measurement represents a direct path from foundational principles to biological practice.

### 3.6.1 A mathematical theory of individuality

A forthcoming work will extend the dilation framework to yield a mathematically explicit theory of biological individuality. In that setting, the complex-valued dilation ratios observed in Figure 10 organize into equivalence classes of informationally coherent paths, grouping lineages into unified "composite individuals" along well-defined curves in the attractor. The dilation equation can then be solved for each composite, which themselves form equivalence classes that can be iteratively re-solved, forming higher-order composites. This recursive resolution filters out all composites of any order by mapping them to the real line. Once complete, the filtered real solutions are treated the same as the real solutions in this analysis.

### 3.6.2 Empirical pathways

Every equation in this work is, in principle, solvable with real biological data. Methods already exist for estimating fractal dimensions in biological branching structures (Katz & George, 1985), reconstructing phylogenetic relationships from genomic data (Felsenstein, 2003), and computing information-theoretic quantities on phylogenies (Li, 1997). The challenge is integrating these into a coherent experimental and statistical pipeline capable of meeting stability and convergence requirements.

This integration is nontrivial. It demands high-resolution lineage data, careful statistical method development, and rigorous treatment of uncertainty. Equally important, the results must be interpreted within a new scientific context. The predictions emerging from this framework are not framed in conventional terms, but in a unified geometry of life and motion that we call *biokinematics*.

### 3.6.3 Biokinematics – life in motion

This measurement formalism opens the door to a new branch of study grounded in kinematics – a field we call biokinematics. If time is multifractal and lineage-dependent, then motion – though still defined as derivatives of position in classical space–time – must also be interpreted in observer-specific terms. Preliminary extensions suggest that even basic kinematic quantities such as velocity may inherit lineage-dependence, raising empirical questions about how motion is differently decomposed across biological forms.

In the limit where biological differences vanish, the dilation factor becomes unity for all pair ($\partial t_j^i = 1$s), and the framework reduces to the ultrametric structure tacitly assumed in current science. Standard kinematics is thus recovered as a special case of the more general, biologically grounded formalism.

Future work will extend this dilation-based approach to systematically examine velocity, acceleration, and related quantities across biological forms, emphasizing their empirical grounding in lineage-specific durations. This perspective shows that the geometry of life generates tractable new questions about motion, suggesting possible bridges between biological and physical descriptions of dynamics.

We close with a reflection on what time as a biological observable means for science, and for the act of measurement itself.

# 4 Conclusion

No scientific theory can escape its biological conditions of validity, because no scientific theory can escape the fact that measurement is a biological act. Philosophers may debate whether non-biological measurement is possible; but in practice, all measurement in science is biological.

This work confronted that fact directly, developing a mathematically precise framework in which observation and measurement are defined through the informational structure of biological lineages. Building on a derivation of time from within a multifractal phylogeny (Hudnall & D'Souza, 2025), the observer was treated not as an abstract agent but as a lineage-dependent entity embedded in the tree of life. Observation, information, and dynamics are unified: the act of measurement is grounded in the structure of the measurer, and the observer becomes a product of the system it observes.

At the technical core is a lineage-dependent dilation equation (Equation 18), which expresses observer-relative time elapse as dimensionless ratios that tie operational standards, such as the SI second, to the biological lineage that defines them. Time remains recorded in seconds, but those seconds are biologically anchored, enabling cross-species conversion through the shared information geometry of descent. This renders the framework empirically tractable: dilation ratios can be estimated from phylogenetic and information-theoretic data, allowing comparative analyses of duration and kinematics across lineages. This connection ensures that the formalism is not abstract but directly measurable: the dilation equation yields ratios that can be recovered in fully operationalized units such as the SI second, linking the mathematics to biological observation.

In this expanded view, the geometry of life, the flow of information, and the act of measurement become one system. Time, observation, and measurement are not primitive background assumptions, but consequences of life's structure. Rather than contradicting physical theory, this framework extends it into the biological domain, where observation, form, and descent are inseparable. Nothing is broken – only expanded. The foundation is deepened by the recognition that observation is not external, but alive.

# Declaration of generative AI and AI-assisted technologies in the writing process

# Declaration of Competing Interest Statement

The author declares no known competing personal or financial interests that could have appeared to influence the work reported in this paper.

# Acknowledgments

I thank Raissa D'Souza for her guidance and mentorship throughout my PhD, of which this work forms a chapter. I am especially grateful for her early recognition of the scientific merit of this research and her continual support as it developed. I thank James Crutchfield for teaching me information theory in Spring 2021, for generously answering questions as they arose during the course of this work, and for providing valuable feedback on its ideas. His recognition that the core issue lay in measurement was especially helpful. I also thank Bryan Jenkins for his careful reading of earlier drafts and his thoughtful responses. I am grateful to Tina Jeoh and Shrinivasa Upadhyaya for believing in this research well before there were results to show. Their support ensured that this research could proceed. Finally, I thank the entire Biological and Agricultural Engineering Department at UC Davis for the opportunity to pursue this research as a PhD. This research received no external funding.

# Appendix A: Three foundational principles of life

The model formalizes three principles: nestedness, duality, and randomness (Hudnall & D'Souza, 2025).

## A.1: The nested principle

The nested principle states that phylogenetic space is convex: the support of a descendant's distribution is a proper subset of its ancestor's, as the ancestor is prerequisite for the descendant. In a branching tree, these nested supports form unique convex hulls (Equation 1) – intersections of ancestor-defined supports – yielding directional lineages as sequences of convexly related distributions.

Mapping the tree to the $n$-dimensional unit interval ($n$ = number of reproduction events) separates phylogenetic space into scale (biological form) and branch length (lifespan). Given the ancestor of all life equals one with certainty (Expression 4), each path defines a unique convex hull in probability space. Duality and randomness ensure path uniqueness.

## A.2: The duality principle

The duality principle holds that the tree of life is shaped by two processes: fission (singularities → populations) and fusion (populations → singularities). Here we model only fission; incorporating fusion would require a network rather than a tree (Section 2.2.1). In our iterated function system (Appendix B), duality is implemented via the Galton-Watson branching process (Equation B.2).

## A.3: The random principle

The random principle states that every path is a stochastic process. The tree is a probability space:

**Expression A1**

$$\left(\mathbb{L}^{D_T=0} \cup \mathbb{T}^{D_T=1}, \mathbb{T}_n^{\text{Dead}}, Pr\right),$$

where $\mathbb{L}^{D_T=0}$ are leaves (dimension 0), $\mathbb{T}^{D_T=1}$ are trees (dimension 1), $\mathbb{T}_n^{\text{Dead}}$ is the σ-field sequence of realized events across reproductions $n$, and $Pr$ is the probability measure.

A random variable $B_l$ is static if, for $n < N$, it equals some event $\beta_l$ with probability 1, belonging to **the dead tree of life**:

**Definition A1**

$$\mathbb{T}_n^{\text{Dead}} = \{\text{All } B_{:i}(n,\cdot) \mid \exists\, \beta_{:i} \text{ with } Pr[B_{:i} = \beta_{:i}] = 1\}.$$

It is dynamic if, at $n = N$, it has nonzero probability of a future form, belonging to **the living tree of life**:

**Definition A2**

$$\mathbb{T}_N^{\text{Alive}} = \left\{\text{All } B_{:k_l}(N, \beta_{:k}) \mid \exists\, \beta_{:i} \text{ with } Pr\left(B_{:k_l} \leq \beta_{:i} < \beta_{:k}\right) \neq 0 \text{ or } 1, \quad \beta_{:k} \in \mathbb{T}_{n<N}^{\text{Dead}}\right\}.$$

Living variables have $(\partial B / \partial t)_l \neq 0$ (Section 2.3). Each living random variable $B_m$ has an associated distribution function:

**Equation A2**

$$F_{:m}(x) = Pr(B_{:m} \leq x), \qquad x \in supp(B_{:m}),$$

with the child's distribution dependent on its parent.

### A.3.1: Model implementation

We model neutral drift using uniform distributions: $B_m \sim U[0, A(B_m))$, with $A(B_m)$ the form of its most recent ancestor. The root ancestor is set to 1. This gives

**Equation A3**

$$F_{:l_m}(x) = P\left(B_{:l_m} \leq x\right) = \frac{x}{\beta_{:l}}, \qquad x \in [0, \beta_{:l}).$$

With density (Equation 5, repeated):

$$f_{:l_m}(x) = \frac{1}{\beta_{:l}}, \qquad x \in [0, \beta_{:l}).$$

These define a nested, monotonically decreasing chain of supports, forming a nonstationary, directed Markov chain:

**Expression A3**

$$P\left(B_{0_{:k_{l_m}}} \leq x \,\middle|\, B_{0_{:k_l}} = \beta_{0_{:k_l}}, B_{0_{:k}} = \beta_{0_{:k}}, \ldots, B_0 = \beta_0\right) = P\left(B_{0_{:k_{l_m}}} \leq x \,\middle|\, B_{0_{:k_l}} = \beta_{0_{:k_l}}\right).$$

Each living entity is the limit (Appendix D) of such a chain.

# Appendix B: The random iterated function system

We generate the simulated trees using a *branching-process-valued* random iterated function system first introduced by (Hudnall & D'Souza, 2025). The iterated map is a Galton-Watson branching process applied at a randomly chosen scale $\beta \in (0,1)$ strictly smaller than the parent's. Iterating this process maps the tree of life to the $n$-dimensional unit interval, where $n$ is the number of reproduction events.

This construction captures the three principles of Appendix A: nestedness (child supports are contained within parent supports), duality (each iteration replaces a point with a tree), and randomness (both scale factors and branching outcomes are stochastic).

### B.1: Classical random iterated function systems

A classical random iterated function system consists of a finite family of maps on a complete metric space, with selection probabilities, and each map typically assumed to be a contraction (Barnsley, 1988; Hutchinson, 1981). The Hutchinson operator acts on compact sets in the Hausdorff metric and, under contractivity, converges to a unique random attractor.

### B.2 Generalization to branching-process-valued maps

Here each map generates a finite rooted subtree via a Galton-Watson process and embeds it in $[0,1]^n$. For each leaf $i$ with scale $\beta_i$, the branching map is defined as:

**Equation B1**
$$T_i^K = \{Z\} \star \beta_i = \mathcal{T}_i \in \mathbb{T}(X),$$

where $\{Z\}$ is Galton-Watson realization:

**Equation B2**
$$Z_{p+1} = \begin{cases} \xi_1^{p+1} + \cdots + \xi_{Z_p}^{p+1}, & Z_p > 0 \\ 0, & Z_p = 0 \end{cases}, \quad 0 < p < G, \quad 0 \leq r < M,$$

with i.i.d. offspring counts $\xi_r^p \in \mathbb{Z}_{\geq 0}$, and finite bounds on generations $G$ and offspring per node $M$ (in simulations $G = 2$ and $M = 3$).

All leaves in a generated subtree share the same $\beta$ in this implementation (Section 2.2.1), yielding a ratio list $\{\beta_{i_1}, \beta_{i_2}, \ldots, \beta_{i_{L(\mathcal{T}_i)}}\}$ and corresponding function list $\{T_{i_1}^K, T_{i_2}^K, \ldots, T_{i_{L(\mathcal{T}_i)}}^K\}$. Here each $\beta_{i_j}$ is drawn from the nested interval distribution (Expression 4), and $L(\mathcal{T}_i)$ gives the number of leaves in subtree $\mathcal{T}_i$. Each $T_{i_j}^K$ produces a Galton-Watson subtree and uniformly contracts it by $\beta_{i_j}$ about the position of the parent.

### B.3: Contraction-map formulation

Although the maps generate finite trees rather than directly mapping $X$ into itself, the construction remains compatible with the contraction theory by working in the hyperspace of compact sets and applying set-valued contractions. Let $\mathcal{K}(X)$ be the space of nonempty compacts in $X$ with the Hausdorff metric $d_H$. Each embedded finite tree $\mathcal{T} \in \mathbb{T}(X)$ is compact, so $\mathbb{T}(X) \subset \mathcal{K}(X)$.

Define a set-valued Hutchinson-Nadler operator (Nadler, 1969) that, given a compact set $K$ representing the current forest, produces the union of all child subtrees generated at scales $\beta \leq \beta_* < 1$:

**Equation B3**
$$\mathcal{H}(K) = \bigcup_{i \in L(K)} T_i(\beta_i)(\{i\}).$$

Each update has a geometric part – a uniform contraction by at most $\beta_* < 1$ in $X$ – and a combinatorial part (finite branching) that does not increase geometric diameter and remains within the contracted support. Under these conditions, standard results for set-valued contractions (Hutchinson, 1981; Nadler, 1969) ensure that $\mathcal{H}$ is contractive on $\mathcal{K}(X)$ and admits a unique random attractor in distribution.

This situates the present branching-process random iterated function system firmly within the established fixed-point theory, with the novelty that each "map" is a topology-changing, branching-process generator followed by a geometric contraction.

### B.4: Output structure

Iteration produces a multifractal "fractal of finite trees," where each lineage is a path of nested random variables supported on shrinking subintervals of $[0,1]^n$. Changing scale induces fission, switching between leaves (dimension 0) and trees (dimension 1). The invariant set of the attractor is the singleton $\{0\}$ (Appendix D.1).

### B.5 Metric versus nearness (construction vs. biology)

The random iterated function system is defined on a complete metric space, so its limit set inherits a Euclidean metric despite its multifractal geometry. In biological terms, however, distance is determined by ancestry and lineage-relative scaling, yielding a nearness relation rather than a global metric (Appendix F). Centering on an observer equips the local neighborhood with a measure, converting dimensionless nearness ratios into operational durations without assuming a background temporal metric.

### B.6 Additional note – relaxing the "unique leaves" assumption

Allowing leaves within a subtree to have different scales increases multifractality but leaves biological interpretation unchanged, since paths remain unique except at the final generation; leaves in a terminal subtree can be grouped without loss.

## Appendix C: Multifractality

As shown in (Hudnall & D'Souza, 2025), each path from the iterated function system is a distinct monofractal so that the ensemble is multifractal.

### C.1.1 Multifractality confirmed in simulation results

We applied multifractal detrended fluctuation analysis (Kantelhardt et al., 2002) to simulated paths: cumulative sum of deviations from the mean, segmentation (scales $3-7$), linear detrending, and calculation of fluctuation functions $V_q(n)$ for $q \in \{-5, -3, -2, -1, 0, 1, 2, 3, 5\}$. The generalized Hurst exponent – slope of $log\left(V_q(n)\right)$ vs. $log(n)$ – was constant across $q$ within each path but varied between paths.

**Figure C1** gives Hurst exponents and confirms multifractality for the system in Figure 5. Each lineage traces a distinct scaling trajectory through form space, shaped by ancestral constraints and entropy accumulation, producing a full spectrum of fractal dimensions (Figure 6, top panel).

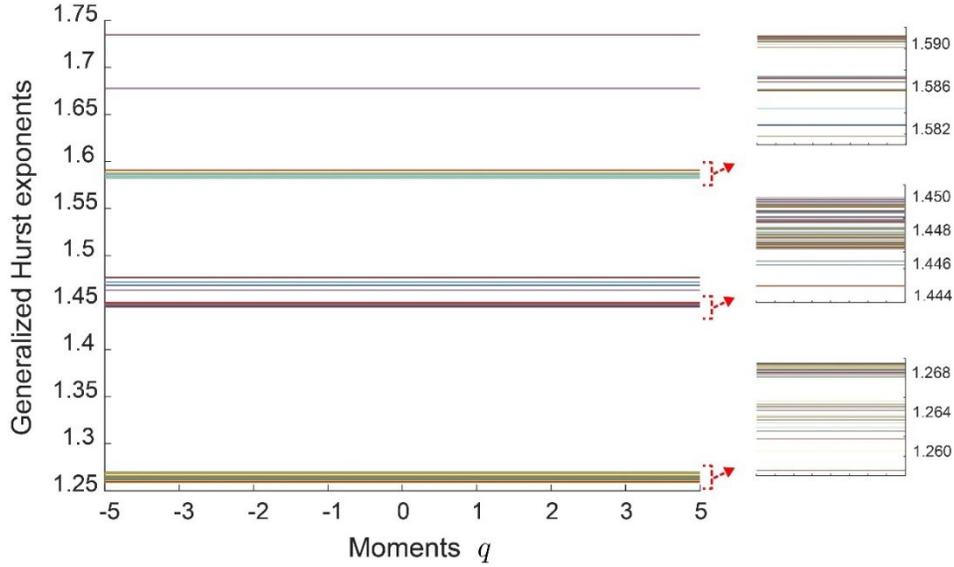

**Figure C1** *The system is multifractal.* Generalized Hurst exponents for 50,000 simulated lineages. Flatness across $q$ shows monofractality within lineages; variation between curves confirms multifractality.

# Appendix D: Convergence – a (Cantor) dust model

A living biological entity is modeled as the limit of a path (Equation 2): a sequence of nested distributions on subintervals of the unit interval converging to a point. As shown in Hudnall & D'Souza (2025), this follows directly from Cantor's nested-set theorem and from convergence in probability: each lineage's support is a decreasing sequence of closed sets with vanishing diameter, or equivalently, the probability of exceeding any fixed $\varepsilon > 0$ tends to zero – biologically, death is certain. Here we show the same result via the invariant set of the random iterated function system, then give the finite-but-large approximation used in simulations and validate the 10% convergence tolerance of Section 2.5.

## D.1: Invariant set of the random iterated function system

Let $X = [0,1]^n$ with the Euclidean metric, and $\mathcal{K}(X)$ the hyperspace of nonempty compact subsets of $X$ with Hausdorff metric $d_H$. Each embedded finite tree $\mathcal{T}$ is in $\mathcal{K}(X)$. For $K \in \mathcal{K}(X)$, define the Hutchinson-Nadler operator (Hutchinson, 1981; Nadler, 1969):

$$\mathcal{H}(K) = \bigcup_{i \in L(K)} T_i(\beta_i)(\{i\}),$$

where $T_i(\beta)$ generates a finite Galton-Watson subtree rooted at $i$ and contracted by $\beta \in (0,1)$ with $\beta_i \leq \beta_* < 1$. Supports are nested with left endpoint at 0 and $B_0 = 1$ a.s

**Lemma D1 (Uniform contraction)**

> For any $K \in \mathcal{K}(X)$, $diam(\mathcal{H}(K)) \leq \beta_* \, diam(K)$.

**Proof**

> The geometric contraction $\beta_* < 1$ bounds the diameter; finite branching cannot enlarge it. ∎

**Proposition D1 (Contractivity)**

$\mathcal{H}: \mathcal{K}(X) \to \mathcal{K}(X)$ is a strict contraction in $d_H$, so there is a unique compact invariant set $A \in \mathcal{K}(X)$ with $\mathcal{H}(A) = A$ (Hutchinson, 1981; Nadler, 1969), and for any $K_0 \in \mathcal{K}(X)$,

$$d_H\big(\mathcal{H}^k(K_0), A\big) \to_{k \to \infty} 0.$$

**Proof**

Follows from Lemma D1 and the completeness of $(\mathcal{K}(X), d_H)$. ∎

**Lemma D2 (Shrinking diameters)**

For any $K_0 \in \mathcal{K}(X)$,

$$diam\big(\mathcal{H}^k(K_0)\big) \leq \beta_*^k \, diam(K_0) \to_{k \to \infty} 0.$$

**Proof**

Follows by iteration of Lemma D1. ∎

**Lemma D3 (Nested supports anchored at 0)**

Along any lineage, the supports $\{I_k\}_{k \geq 0}$ satisfy $I_k \subseteq [0, \beta_{k-1}] \subseteq [0,1]$ with fixed left endpoint 0.

**Proof**

Starting from $supp(B_0) = [0,1]$, each child's support is contained in $[0, \beta_{\text{parent}})$ with the left endpoint inherited unchanged (Appendix A.3.1), ensuring nested supports along the lineage. ∎

**Theorem D1 (The invariant set)**

The unique compact invariant set is $A = \{0\}$.

**Proof**

By Lemma D2, $diam\big(\mathcal{H}^k(K_0)\big) \to 0$. By Lemma D3, the left endpoint is anchored at 0. By Cantor's nested set theorem, the intersection $\bigcap_{k \geq 0} \overline{\mathcal{H}^k(K_0)}$ is the singleton $\{0\}$; uniqueness follows from Proposition D1. ∎

## D.2: Biological realism – the finite-but-large approximation

Mathematically, all lineages converge to $\{0\}$ biologically, paths are finite. For large $n$, an entity is effectively a single form – the random variable on the large-limit intersection of its ancestral supports – with small but nonzero variance until death. This mirrors biology: life approaches the limit without reaching it.

**Lemma D4 (Finite-but-large convergence sketch)**

Under contractivity, each step shrinks by a factor $\rho_k < 1$, so $diam(B_{:n}) \leq C \prod_{k=1}^{n} \rho_k$. In particular, if the $\rho_k$ are uniformly bounded above by some $\rho < 1$, then $diam(B_{:n}) \leq C\rho^n$. Hence for any tolerance $\tau > 0$ there is $n_\tau$ with $|D\_F(n) - D\_F(N)| \leq \tau |D\_F(N)|$ for all $n \geq n_\tau \ll N$.

## D.3: Sensitivity analysis of the convergence tolerance

Let $D_F(n)$ be the correlation dimension (Section 2.4.1) at iteration depth $n$, and $N$ the terminal depth. For $\tau \in \{5,8,10,12,15,20\}\%$, define the **first stable depth** $n_c(\tau)$ as the smallest $n$ with:

**Definition D1**

$$|D_F(m) - D_F(N)| \leq \tau|DF(N)|, \quad \forall\, m \geq n.$$

A path is **persistently converged** at $n$ in the above holds for later depths.

**Figure D.1 LEFT** shows that the relative error decreases monotonically and the interquartile range (IQR, shaded) collapses, with population-level stabilization of $D_F(n)$ before $N$. **Figure D.1 RIGHT** shows that the percentage of converged paths rises sharply in the final iterations, with larger τ shifting convergence earlier, as expected.

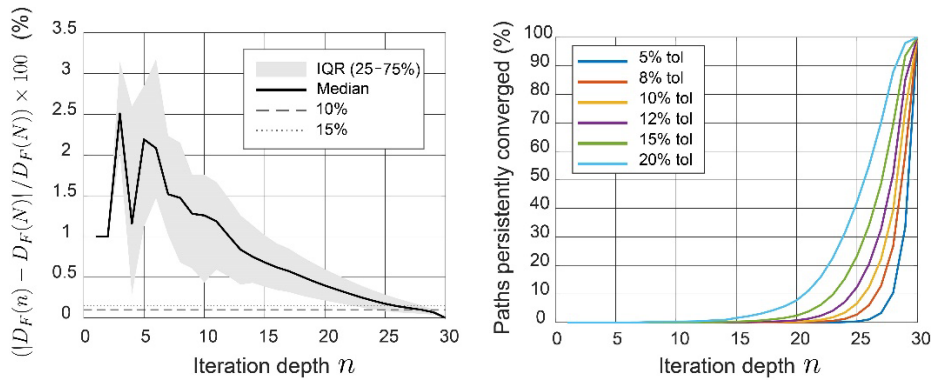

**Figure D1 (LEFT)** Median ± IQR of relative error vs. iteration depth $n$, with reference lines at $\tau = 10\%$ and 15%. **(RIGHT)** Cumulative fraction of paths converged as a function of $n$ for tolerances $\{5,8,10,12,15,20\}\%$.

**Figure D.2** summarizes tolerance dependence. **(A)** shows the fraction of paths converged by $N-2$ iterations rises monotonically with τ. **(B)** shows that the depths for 50%, 80%, and 90% of convergent paths decreases as τ increases. And **(C)** shows that the median first stable depth $n_c(\tau)$ (blue, with IQR band) and the mean endpoint depth $n_e$ (red dashed) both decrease smoothly with τ. For $\tau \in [10,15]\%$, $n_c$ is typically within the final one-two iterations.

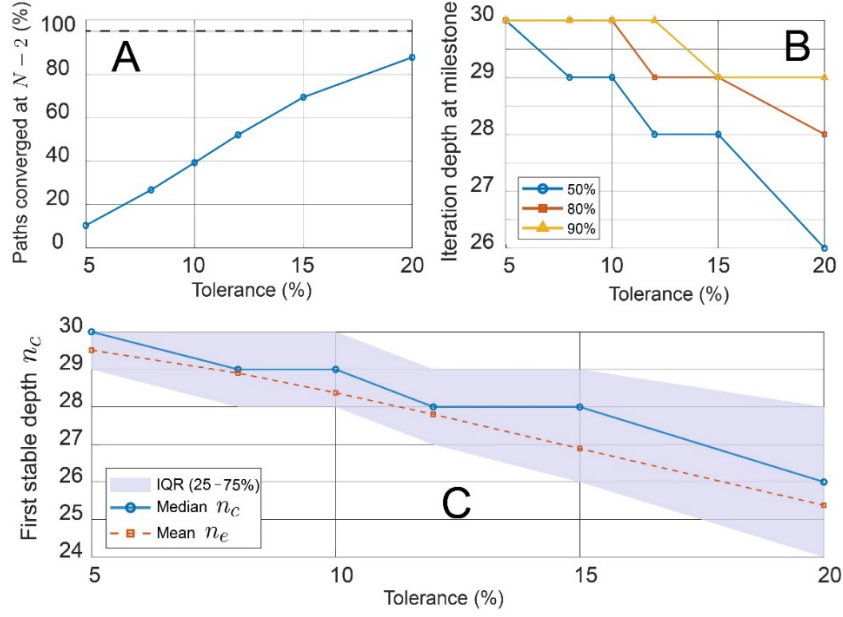

**Figure D2 (A)** Fraction of paths converged by $N-2$ vs. tolerance. **(B)** Iteration depth where 50%, 80%, and 90% of paths are persistently converged vs. tolerance. **(C)** Median (±IQR) first stable depth $n_c$ and mean endpoint depth $n_e$ vs. tolerance.

These results show that the 10% criterion of Section 2.5 is conservative, yielding late-stage but robust acceptance and consistent with the finite-but-large approximation. All convergence-dependent quantities in the main text are stable with respect to reasonable changes in τ.

# Appendix E: The correlation dimension $D_F$

The correlation dimension method (Grassberger & Procaccia, 1983) estimates the fractal dimensionality of each lineage's ancestral form space. For path $B_i$ with $N$ points, the correlation integral is:

**Equation E1**

$$C_i(\epsilon) = \lim_{N \to \infty} \frac{g_i}{N^2},$$

where $g_i$ counts point pairs separated by less than $\epsilon$. Scaling follows:

**Expression E1**

$$C_i(\epsilon) \sim \epsilon^{D_F(B_{:i})},$$

with $D_F(B_{:i})$ given by the slope of the regression line for $log(C(\epsilon))$ against $log(\epsilon)$.

We used 12,000 logarithmically spaced $\epsilon$ values between 0 and 1, omitting endpoints and zeros in $C(\epsilon)$. Linear regression on the retained pairs yields $D_F$.

## E.1: Method validation

**Figure E1** shows method performance.

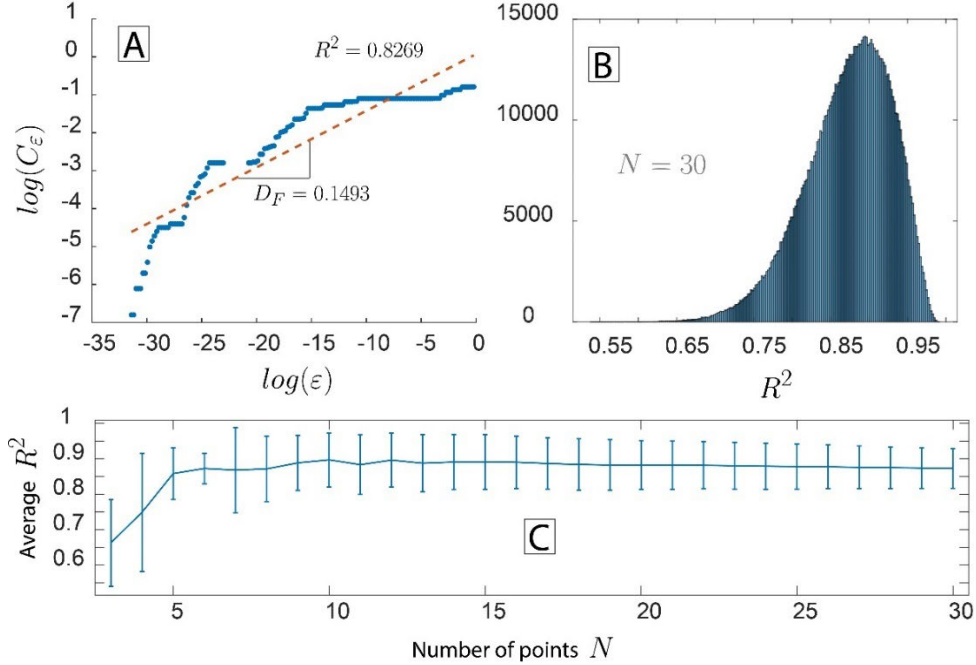

**Figure E1** *Statistics for estimating the correlation dimension.* **(A)** $D_F$ for a randomly chosen path, computed using 143 values of $\varepsilon$ ranging from 0.81 to $1.23e^{-12}$. **(B)** Histogram of $R^2$ values across all paths at iteration 30; mean $R^2 = 0.869 \pm 0.0499$. **(C)** Mean $R^2$ at each iteration with standard deviation error bars. Stability is reached by iteration 5.

# Appendix F: Life is a nearness space, not a metric space

The multifractal structure generated by the random iterated function system (Appendix B) lacks globally fixed distances: values depend on the observer's position via local scaling exponents. Without fixing an observer, the "distance" between two points varies by comparison path, making it neither single-valued (not well-defined) nor consistent with the metric axioms. Distinct points can have zero separation under one scaling exponent and nonzero under another, and triangle inequalities may fail when paths traverse regions with differing exponents.

## F.1 Nearness space formalism

We consider the whole space – the set of all leaves ($\mathbb{L}^{D_T=0}$) and subtrees ($\mathbb{T}^{D_T=1}$) of topological dimensions zero and one respectively:

**Expression F1**

$$\mathbb{X} = \mathbb{L}^{D_T=0} \cup \mathbb{T}^{D_T=1}.$$

Instead of a metric $d: \mathbb{X} \times \mathbb{X} \to \mathbb{R}$, the tree admits a **nearness relation** $\partial t_j^i$ (Equation 18) on nonempty subsets of $\mathbb{X}$, where $\partial t_j^i \neq 0$ if and only if $i$ and $j$ share a common ancestor. This relation, given by the solution to the dilation equation, holds for both real (pure divergence) and complex (coherence) solutions (Sections 3.3.3–3.4). This makes every object near every object (except by the origin where $\partial t_j^i$ is not well-defined) which is necessary since the system is unified by common descent.

## F.2.1 Graded nearness

In this formalism, the argument $arg(\partial t^i_j)$ encodes the qualitative nature of the relationship: $arg(\partial t^i_j) = 0$ implies pure divergence (real solution); $arg(\partial t^i_j) \neq 0$ implies coherence (complex solutions) (Section 3.4). When $\partial t^i_j$ is real, the magnitude gives "how near" $i$ is to $j$ – ultimately how similar their experienced durations are. When $\partial t^i_j$ is complex, magnitudes and arguments organize such dilations into an attractor on the complex plane (Figure 10, Section 3.4).

The nearness relation emerges from the evolutionary dynamics itself, replacing fixed distance with observer-relative nearness shaped by each lineage's branching and contraction history.

## F.2.2 Proof of nearness relation

For $i, j \in \mathbb{X}$ with $i \neq j$, let the information theoretic dilation $\partial t^i_j$ be defined (most recent common ancestor exists greater than two generations back; convergence requirements are met – section 2.5). Set $\partial t^i_i = 1$ (centering, Section 2.7.4), and let $\text{Dil}(i, j)$ mean "$\partial t^i_j$ is well-defined".

**Definition F1 (Nearness on nonempty subsets)**

> Let $\mathcal{P}^*(\mathbb{X})$ be the family of nonempty subsets of $\mathbb{X}$, with $A, B \in \mathcal{P}^*(\mathbb{X})$. Let $\delta$ be a binary relation on $\mathcal{P}^*(\mathbb{X})$ defined by:
>
> $$A \delta B \iff \exists i \in A, \exists j \in B \text{ such that } \text{Dil}(i, j).$$
>
> (Equivalently, for singletons $\{i\} \delta \{j\}$ iff $\partial t^i_j$ is defined).

**Claim**

> $(\mathbb{X}, \delta)$ is a nearness space; in particular, $\delta$ satisfies the standard nearness axioms (Herrlich, 1974):
>
> **(N1) Nontriviality**: if $A \delta B$ then $A \neq \emptyset$ and $B \neq \emptyset$.
>
> **(N2) Symmetry**: $A \delta B \iff B \delta A$.
>
> **(N3) Union additivity**: $A \delta (B \cup C) \iff (A \delta B) \text{ or } (A \delta C)$.
>
> **(N4) Intersection implies nearness**: if $A \cap B \neq \emptyset$ then $A \delta B$.

**Proof**

> **(N1)** Holds by definition: $i \in A$ and $j \in B$ must exist.
>
> **(N2)** if $A \delta B$, there are $i \in A$ and $j \in B$ with $\text{Dil}(i, j)$. Since the dilation is defined for the pair $(i, j)$, it is also defined for $(j, i)$ (the two directions are both well-defined via the same ancestral data), hence $B \delta A$.
>
> **(N3)** $A \delta (B \cup C)$ means there are $i \in A$ and $x \in (B \cup C)$ with $\text{Dil}(i, x)$. Then either $x \in B$ (giving $A \delta B$) or $x \in C$ (giving $A \delta C$). The converse immediate by taking the same witness in $B \cup C$.
>
> **(N4)** if $x \in (A \cap B)$, then $\partial t^x_x \equiv 1$ by centering, so $\text{Dil}(x, x)$ holds and hence $A \delta B$. ∎

In the ideal, unfiltered theory, any two leaves share a most recent common ancestor, so $\mathrm{Dil}(i,j)$ holds for all $i, j \in \mathbb{X}$. This defines an indiscrete nearness relation, reflecting the biological fact of universal common descent. Finer structure is captured by the complex dilations $\partial t_j^i$, whose magnitudes encode degree of nearness and whose arguments partition solutions into composite-equivalence classes (Section 3.4). In computational practice, convergence filters (Section 2.5.1) restrict attention to stable comparisons, yielding a subrelation of δ; the theoretical relation remains well-defined regardless of these filters.

### F.3 Centering and the emergence of an observer–relative time measure

A well-defined temporal measure is obtained after specifying $p \in \mathbb{X}$ as the center (Section 2.7.4), fixing the dilation equation so that $\partial t_p^p = 1$, and thereby defining the unit of time in $p$'s frame (Section 3.5). The centered dilations $\partial t_p^i$ ($p$ fixed) then express the ratio of elapsed duration experienced by $i$ relative to $p$ in units defined by $p$, defining a measure:

**Expression C3**

$$\mu_p \colon \mathbb{X} \to \mathbb{R}_{>0}, \qquad \mu_p(i) = \partial t_p^i \cdot \partial t_p^p,$$

and the **time-measured space** $\left(\mathbb{X}, \mathbb{T}^{\mathrm{Alive}}, \mu_p\right)$.

### F.4 Summary statement

Externally, the tree's geometry is multifractal and fails to admit a metric; internally, the nearness formalism captures its relational structure, and centering induces a well-defined, observer-relative temporal measure.

# Appendix G: Statistical confirmations of simulation results

Simulation outcomes were validated through repeated trials to ensure statistical robustness. Ten independent evolutionary simulations (30 iterations each) were performed, with 50 million random pairwise comparisons sampled per system.

### G.1: Convergence statistic for crosspath dimensions (Section 3.2.2)

In the system of 976,230 leaves, ~477 billion pairwise comparisons are possible. Of these, 4,551,026 (~0.001%) met convergence criteria (Section 2.5.1). Across trials, the mean convergence rate was 0.003%, with a standard deviation of 0.006%.

### G.2: Coherence-divergence statistics for information distance (Section 3.3.3)

Most comparisons were coherent ($d_\beta > 0$): $5.6015 \times 10^{11}$ (about 99.4%) vs. $3.1739 \times 10^9$ (about 0.6%) divergent ($d_\beta < 0$). Across trials, mean proportions were 94.8% (standard deviation: 12%) for coherent, and 5.2% (standard deviation: 12%) for divergent.

### G.3: Indeterminant comparisons (Section 3.4)

If the most recent common ancestor occurs within three generations, comparisons are classified as indeterminant. In the full system, these numbered 350,727 ($< 0.0001\%$). Across trials: 0.0002% (standard deviation: 0.0003%).

## G.4: Convergence filtering statistics for the dilation equation (Section 3.4)

The dilation equation yields both real-valued and complex-valued solutions, depending on the evolutionary relationship between lineages.

### G.4.1: Complex versus real solutions

Of convergent cases, 4,459,514 solutions (~99.4%) were real-valued (divergent), and 91,512 (~0.6%) were complex-valued (coherent). Across trials: 97.82% (standard deviation: 4.68%) real, 2.18% (standard deviation: 4.68%) complex.

### G.4.2: Convergence for complex versus real solutions

The vast majority of comparisons that are lost due to convergence indeterminacy were complex: $5.6015 \times 10^{11}$ comparisons showed $d_\beta > 0$, but only 91,512 comparisons met the convergence requirement (less than 0.0001%). A far greater percentage of real only solutions were retained: $3.1739 \times 10^9$ comparisons showed $d_\beta < 0$, and 4,459,514 of these met the convergence requirement (about 0.14%). Across trials: 0.0046% (standard deviation: 0.0096%) convergent, 0.15% (standard deviation: 0.10%) divergent.

## G.5: Distribution of observational capacity

The number of target leaves observable per leaf is highly skewed: most observe hundreds to thousands, while a minority observe tens of thousands. The heavy-tailed pattern reflects heterogeneity in local scaling exponents, as expected in a multifractal structure. **Figure G1** shows the resulting histogram on a logarithmic vertical scale.

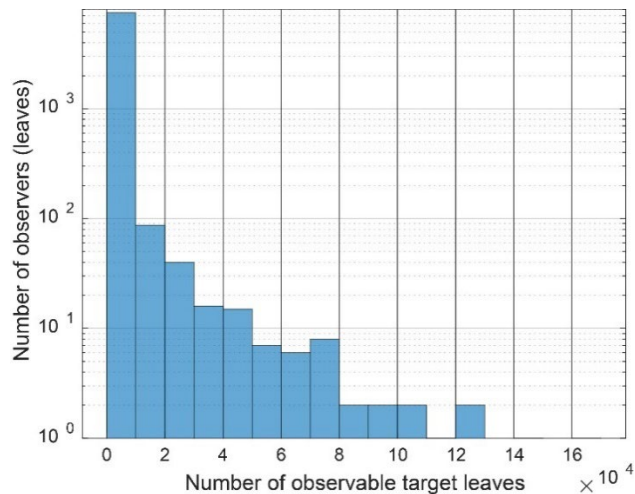

**Figure G1** *Histogram of the number of observable targe leaves per observer leaf.*

# Appendix H: Notation reference table

For ease of reference, **Table H1** consolidates symbols appearing in the main text and appendices.

**Table H1** *Summary of notation used in the manuscript.*

| Symbol | Meaning/Definition | Text reference |
|---|---|---|

| Symbol | Description | Reference |
|---|---|---|
| $\beta_{0:k_l}$; $\beta_l$ | Biological event at lineage step $l$; shorthand version | Section 2.1 |
| $B_{0:l_m}$; $B_m$ | Biological random variable at lineage step $m$; shorthand | Section 2.1 |
| $A(B_{:l_m})$; $A(m)$ | Ancestor of $B_{:l_m}$; shorthand using index for r.v. | Section 2.1 |
| $\mathbb{T}_n^{\text{Dead}}$ | The dead tree of life | Section 2.1 |
| $conv(B_{:m})$ | The convex hull of $B_{:m}$ | Section 2.1 |
| $\partial B_m / \partial t_m$; $(\partial B/\partial t)_m$ | Lineage dynamic; equivalent shorthand | Section 2.3 |
| $\mathbb{T}^{\text{Alive}}$ | The living tree of life | Section 2.3.2 |
| $\lim_{n \gg 1}$ | The finite-but-large reproduction limit | Section 2.4 |
| $D_F(B_i)$; $D_F(i)$ | Fractal dimension of path $i$; shorthand | Section 2.4.1 |
| $D_F(B_i, B_j)$; $D_F(i,j)$ | Crosspath fractal dimension of $i$ and $j$ | Section 2.4.2 |
| $D_T$ | Topological dimension | Section 2.2 |
| $Pr$ | Probability measure | Appendix A.3 |
| $U$ | The uniform distribution | Section 2.6.1 |
| $f_{:l_m}(\cdot)$ | Probability density function for r.v. $B_{:l_m}$ | Section 2.6.1 |
| $H[B_{:m}]$ | Differential Shannon entropy of path $m$ | Section 2.6.1 |
| $F(B_{:l_m})$; $F_{:l_m}(\cdot)$ | Probability distribution function; equivalent | Appendix A.3 |

| Symbol | Description | Reference |
|---|---|---|
| $F(B_i, B_j)$ | Joint probability distribution function | Section 2.6.2 |
| $f_{ij}(\cdot,\cdot)$ | Joint probability density function for r.v.s $B_i$ and $B_j$ | Section 2.6.2 |
| $H\left[B_{\cdot i}, B_{\cdot j}\right]$ | Joint entropy of $i$ and $j$ | Section 2.6.2 |
| $I\left[B_{\cdot i}; B_{\cdot j}\right]$ | Mutual information of $i$ and $j$ | Section 2.6.2 |
| $d_\beta\left(B_{\cdot i}, B_{\cdot j}\right)$ | Information distance between $i$ and $j$ | Section 2.6.2 |
| $\partial t^i_j$ | Temporal dilation ratio/nearness ratio for $i$ and $j$ | Section 2.7.1 |
| $\mathbb{L}^{D_T=0}$ | Leaves of topological dimension zero | Appendix A.3 |
| $\mathbb{T}^{D_T=1}$ | (sub)trees of topological dimension one | Appendix A.3 |
| $T_i^K$ | Function iterated in the RIFS – generates a Galton-Watson tree at iteration $K$ and scale $\beta_i$ | Appendix B.2 |
| $\mathcal{T}_i$ | The scaled Galton-Watson tree generated by $T_i^K$ | Appendix B.2 |
| $\{Z\}$ | The Galton-Watson process | Appendix B.2 |
| $\xi_r^p$ | Offspring count $r$ at generation $p$ for the Galton-Watson process | Appendix B.2 |
| $G$ | Maximum generations in the Galton-Watson process | Appendix B.2 |
| $M$ | Maximum offspring per node in the Galton-Watson process | Appendix B.2 |
| $L(\mathcal{T}_i)$ | The number of leaves in subtree $\mathcal{T}_i$ | Appendix B.2 |
| $X$ | Ambient space of the RIFS | Appendix B.3 |

| Symbol | Description | Reference |
|---|---|---|
| $\mathcal{K}(X)$ | The space of nonempty compact subsets on $X$ | Appendix B.3 |
| $d_H$ | The Hausdorff metric | Appendix B.3 |
| $\mathcal{H}(K)$ | Union of child subtrees | Appendix B.3 |
| $q$ | Moment order parameter; exponent in multifractal analysis where expectations take the form $\sum_i p_i^q$ | Appendix C |
| $V_q(n)$ | Fluctuation function for detrended fluctuation analysis | Appendix C |
| $diam(\cdot)$ | Set diameter | Appendix D |
| $\tau$ | Convergence tolerance value | Appendix D.2 |
| $\rho$ | Branching ratio / probability of lineage bifurcation | Appendix D.2 |
| $C_i(\epsilon)$ | Correlation integral for correlation dimension | Appendix E |
| $\sim$ | Asymptotic scaling relation | Appendix E |
| $R^2$ | Coefficient of determination in regression analysis | Appendix E.1 |
| $\mathbb{X}$ | Union of the set of all leaves ($\mathbb{L}^{DT=0}$) and subtrees ($\mathbb{T}^{DT=1}$); the entire space generated by the RIFS | Appendix F |
| $arg(\partial t_j^i)$ | Argument (phase angle) of the dilation equation solution $\partial t_j^i$ | Appendix F.1 |
| $\mu_p$ | Measure defined by centering on $p$ | Appendix F.3 |
| $\equiv$ | Identity (defined equality) | Throughout |
| $\Delta t_i$ | $i$'s experienced duration | Throughout |

## Data availability

The data that support the findings of this study are openly available in the git repository at https://github.com/KevinAndrewHudnall/Information-and-the-living-tree-of-life.

# Silhouette credits

All silhouettes were accessed at phylopic.org.

Homo habilis By T. Michael Keesey. License: CC0 1.0 Universal Public Domain Dedication.
Biarmosuchus tener By T. Michael Keesey. License: Attribution 3.0 Unported.